\documentclass[english,aps,preprintnumbers,amsmath,amssymb,nofootinbib,twocolumn,10pt,prd]{revtex4-1} 

\usepackage[utf8]{inputenc}
\usepackage[T1]{fontenc}
\usepackage[colorlinks]{hyperref}

\usepackage{ifthen}
\usepackage{forloop}
\usepackage{babel}
\usepackage{graphicx}
\usepackage{psfrag}
\usepackage{color}
\usepackage{dsfont}
\usepackage{mathrsfs}
\usepackage{mathtools} 

\usepackage{slashed}

\usepackage{mathbbol}

\usepackage{ulem}

\usepackage{color}
\definecolor{Turquois}{rgb}{0.00, 0.70, 0.70}
\definecolor{DarkGreen}{rgb}{0.00, 0.60, 0.00}

\newcommand{\parametrization}{parametrization}
\newcommand{\parametrizations}{parametrizations}

\newcommand{\GN}{G_{\text{N}}}

\newcommand{\metric}{\ensuremath{g}}
\newcommand{\flucmet}{\ensuremath{h}}

\newcounter{pointcounter}
\newcommand{\point}[1][\empty]{
  \ifthenelse
    {\equal{#1}{\empty}}
    {\ensuremath{\, \phantom{\cdot} \,}}
    {\setcounter{pointcounter}{1} \forloop{pointcounter}{0}{\value{pointcounter} < #1}{\ensuremath{\, \phantom{\cdot} \,}}}
}

\newcommand{\Eqref}[1]{Eq.~\eqref{#1}}

\newcommand{\Tr}{\operatorname{Tr}}

\newcommand{\mcF}{\ensuremath{\mathcal{F}}}

\newcommand{\mcL}{\ensuremath{\mathcal{L}}}
\newcommand{\mcM}{\ensuremath{\mathcal{M}}}

\newcommand{\mcO}{\ensuremath{\mathcal{O}}}
\newcommand{\mcP}{\ensuremath{\mathcal{P}}}
\newcommand{\mcR}{\ensuremath{\mathcal{R}}}
\newcommand{\mcS}{\ensuremath{\mathcal{S}}}

\newcommand{\mcZ}{\ensuremath{\mathcal{Z}}}

\newcommand{\mrT}{\ensuremath{\mathrm{T}}}

\makeatother

\newcommand{\UV}{{\small UV}}
\newcommand{\IR}{{\small IR}}
\newcommand{\FRG}{{\small FRG}}
\newcommand{\RG}{{\small RG}}

\graphicspath{{./figures/}}

\begin{document}
\allowdisplaybreaks

\title{Generalized Parametrization Dependence in Quantum Gravity
}
\author{Holger Gies, Benjamin Knorr and Stefan Lippoldt}
\affiliation{Theoretisch-Physikalisches Institut, Friedrich-Schiller-Universit\"at Jena, 
Max-Wien-Platz 1, D-07743 Jena, Germany}                                        

\begin{abstract}
We critically examine the gauge, and field-parametrization dependence
of renormalization group flows in the vicinity of non-Gau\ss{}ian
fixed points in quantum gravity. While physical observables are
independent of such calculational specifications, the construction of
quantum gravity field theories typically relies on off-shell
quantities such as $\beta$ functions and generating functionals and
thus face potential stability issues with regard to such generalized
\parametrizations. We analyze a two-parameter class of covariant gauge
conditions, the role of momentum-dependent field rescalings and a
class of field parametrizations. Using the product of
Newton and cosmological constant as an indicator, the principle of
minimum sensitivity identifies stationary points in this
\parametrization\ space which show a remarkable insensitivity to the
\parametrization. In the most insensitive cases, the quantized gravity
system exhibits a non-Gau\ss{}ian \UV{} stable fixed point, lending
further support to asymptotically free quantum gravity. One of the
stationary points facilitates an analytical determination of the
quantum gravity phase diagram and features ultraviolet and infrared
complete \RG{} trajectories with a classical regime.
\end{abstract}

\maketitle

\section{Introduction}\label{sec:intro}

Physical observables are independent of their computational
derivation. Still, many practical computations are based on convenient
choices for intermediate auxiliary tools such as coordinate systems,
gauges, etc.  Appropriate \parametrizations\ of the details of a system
simply decrease the computational effort. Beyond pure efficiency
aspects, such suitable \parametrizations\ can also be conceptually advantageous
or even offer physical insight. This is similar to coordinate choices
in classical mechanics where polar coordinates with respect to the
ecliptic plane in celestial mechanics support a better understanding
in comparison with, say, Cartesian coordinates with a $z$ axis
pointing towards Betelgeuse.

Appropriate parameterizations become particularly significant in
quantum calculations. While on-shell quantities such as $S$-matrix
elements are invariant observables
\cite{Coleman:1969sm,Kallosh:1972ap,'tHooft:1973pz}, off-shell
quantities generically feature \parametrization\ dependencies,
including gauge-, field-parametrization and regularization-scheme
dependencies
\cite{ZinnJustin:2002ru,Jackiw:1974cv,Nielsen:1975fs}. Further
ordering schemes such as perturbative expansions may defer such
dependencies to higher orders (such as scheme dependence in
mass-independent schemes), but these are merely special and not always
useful limits. Approximation schemes that can also deal with
non-perturbative regimes may even introduce further artificial
\parametrization\ dependencies which have to be carefully removed
(e.g., discretization artefacts in lattice regularizations).

In an ideal situation, this \parametrization\ dependence of a
nonperturbative approximation could be quantified and proven to be
smaller than the error of the truncated solution. However, as soon as
a result is \parametrization\ dependent, it is likely that some
pathological \parametrization\ can be constructed that modifies the
result in an arbitrary fashion. This suggests to look for general
criteria of \textit{good \parametrizations} that minimize the
artificial dependence in approximation schemes which adequately
capture the physical mechanisms.

\textit{A-priori} criteria suggest the construction of
\parametrizations\ that support the identification of physically
relevant degrees of freedom, such as the use of Coulomb-Weyl gauge in
quantum optics, or the use of pole-mass regularization schemes in
heavy-quark physics. Further a-priori criteria include symmetry
preserving properties (covariant gauges, non-linear field
parametrizations) or strict implementations of a \parametrization\ condition
such as the Landau-gauge limit $\alpha\to 0$. A major advantage of the
latter is that some redundant degrees of freedom decouple fully from
the dynamical equations in such a limit.

Good \parametrizations\ may also be identified \textit{a posteriori} by allowing
for a family of \parametrizations\ and identifying stationary
points in the parameter space. This realizes the principle of
minimum sensitivity \cite{Stevenson:1981vj,Ball:1994ji} (originally advocated for
regularization-scheme dependencies), suggesting those points as
candidate parameters for minimizing the influence of \parametrization\
dependencies.

In the present work, we investigate a two-parameter family of
covariant gauges, a family of field parametrizations and the role of
momentum-dependent field rescalings in quantum gravity in this
spirit. The family of gauges includes a (non-harmonic) generalization
of the harmonic gauge (De-Donder gauge), the latter being particularly
useful for the analysis of gravitational waves which presumably are
the asymptotic states of quantum gravity. The a-priori criteria
suggest to implement this gauge in the Landau-gauge limit to decouple
a redundant part of the Hilbert space.  In fact, in this limit we find
a subtlety in the form of a degeneracy in the subspace of scalar field
components which is special to gravity.

We also investigate a one-parameter family of field parametrizations
that includes the most widely used linear split \cite{DeWitt:2003pm}
as well as the exponential split \cite{Kawai:1993fq,Kawai:1992np,
  Kawai:1993mb,Kawai:1995ju,Aida:1994zc} studied more recently in the
context of asymptotic safety
\cite{Nink:2014yya,Demmel:2015zfa,Percacci:2015wwa,Labus:2015ska} --
both of which find support by discriminative a-priori arguments. We
also take a brief look at the most general ultra-local four-parameter
family of parametrizations to quadratic order, corroborating the
results of the one-parameter family. In addition, we study the
influence of momentum-dependent field rescalings which are commonly
used in gravity in connection with the York decomposition. In the
context of the functional renormalization group (\FRG{})
\cite{Wetterich:1992yh} which provides a tool to study quantum gravity
nonperturbatively \cite{Reuter:1996cp}, these parametrization
dependencies can mix nontrivially with the regularization of the
spectrum of fluctuations. Therefore, the analysis of parametrization
dependencies also explores implicitly the stability of the system in
the ultraviolet.

Interestingly, we observe a nontrivial interplay between all these
\parametrization\ dependencies. Still, several stationary points can
be observed in the results for the \RG{} flow where the system
develops a remarkable insensitivity to the details of the
\parametrization\ choices. In particular, for the stable
\parametrizations, we observe the existence of a \UV{} stable
non-Gau\ss{}ian fixed point which provides further quantitative
evidence for the existence of an asymptotically safe metric quantum
gravity \cite{Weinberg:1976xy,Weinberg:1980gg}. In the stationary regime
of the \parametrization\ based on the exponential split, the resulting
\RG{} flow exhibits several remarkable properties: (1) a possible
dependence on the residual gauge parameter drops out implying an
enhanced degree of gauge invariance, (2) the \RG{} flow becomes
particularly simple, such that the phase diagram in the plane of
Newton and cosmological constant can be computed analytically, (3) no
singularities arise in the flow, such that a large class of \RG{}
trajectories (including those with a classical regime) can be extended
to arbitrarily high and low scales, (4) the \UV{} critical exponents
are real and close to their canonical counterparts, and (5)
  indications are found that the asymptotic safety scenario may not
  extend straightforwardly to dimensions much higher than $d=4$.

\section{Quantum gravity and \parametrizations}\label{sec:choosing_a_gauge}

The technical goal of quantum gravity is to construct a functional
integral over suitable integration variables which in the long-range
limit can be described by a diffeomorphism-invariant effective field
theory of metric variables approaching a classical regime for a wide
range of macroscopic scales. The fact that the first part of this
statement is rather unspecific is reflected by the large number of
legitimate quantization proposals
\cite{Kiefer:2004gr,Rovelli:2004tv,Ashtekar:2014kba}. Independently of
the precise choice of integration variables, a renormalization group
approach appears useful in order to facilitate a scale-dependent
description of the system and a matching to the long-range classical
limit which is given at least to a good approximation by an
(effective) action of Einstein-Hilbert type:
\begin{equation}
\Gamma_{k} = -\int \text{d}^dx \sqrt{\metric} \, \mcZ_{R} (R \! - \! 2 \Lambda).
\label{eq:EH}
\end{equation}
Here, we have already introduced a momentum scale $k$, expressing the
fact that this effective description should a priori hold only for a
certain range of classical scales. In this regime, we have $\mcZ_{R}=
1/(16\pi \GN)$ with the Newton constant $\GN$, and $\Lambda$
parametrizing the cosmological constant. In a quantum setting, 
$\mcZ_{R}$ plays the role of a (dimensionful) wave-function renormalization, and
$\GN$ and $\Lambda$ are expected to be replaced by their running
counterparts depending on the scale $k$.

In the present work, we confine ourselves to a quantum gravity field
theory assuming that the metric itself is already a suitable
integration variable. A first step towards a diffeomorphism-invariant
functional integral then proceeds via the Faddeev-Popov method
involving a gauge choice for intermediate steps of the calculation. In
this work, we use the background-field gauge with the gauge-fixing
quantity,
\begin{equation}
F_{\mu} = \left( \delta_{\mu}^{\beta} \bar{D}^{\alpha} - \frac{1 + \beta}{{d}} \bar{\metric}^{\alpha \beta} \bar{D}_{\mu} \right) \metric_{\alpha \beta}, 
\end{equation}
which should vanish if the gauge condition is exactly matched. Here,
$\metric_{\alpha\beta}$ is the full (fluctuating) metric, whereas
$\bar{\metric}_{\alpha\beta}$ denotes a fiducial background metric
which remains unspecified, but assists to keep track of diffeomorphism
invariance within the background-field method. Gauge-fixing is
implemented in the functional integral by means of the gauge-fixing
action 
\begin{equation}
\Gamma_{\text{gf}}= \frac{\mcZ_{R}}{2 \alpha} \int {\text{d}}^dx
\sqrt{\bar{\metric}} \bar{\metric}^{\mu \nu} F_{\mu} F_{\nu}.
\label{eq:Sgf}
\end{equation}
More precisely, this gauge choice defines a two-parameter
($\alpha,\beta$) family of covariant gauges. For instance, the choice
$\beta=1$ corresponds to the harmonic/De-Donder gauge which together
with $\alpha=1$ (Feynman gauge) yields a variety of technical
simplifications, being used in standard effective field theory
calculations \cite{Donoghue:1993eb,BjerrumBohr:2002kt,Robinson:2005fj}
as well as in functional \RG{} studies \cite{Reuter:1996cp,Reuter:2012id} of quantum
gravity. More conceptually, the Landau-gauge limit $\alpha\to0$
appears favorable, as it implements the gauge condition in a strict
fashion and thus should be a fixed point under \RG{} evolution
\cite{Ellwanger:1995qf,Litim:1998qi}.

In the Euclidean formulation considered here, the parameter $\alpha$
is bound to be non-negative to ensure the positivity of the
gauge-fixing part of the action (this restriction may not be necessary
for a Lorentzian formulation). The parameter $\beta$ can be chosen
arbitrarily except for the singular value $\beta_{\text{sing}}={d-1}$. To
elucidate this singularity, let us take a closer look at the induced
Faddeev-Popov ghost term:
\begin{equation}
\Gamma_{\text{gh}}=- \int {\text{d}}^dx \sqrt{\bar{\metric}} \bar{C}_{\mu} \mcM^{\mu}_{\point \nu} C^{\nu}, \quad 
 \mcM^{\mu}_{\point \nu} = \frac{\delta F^{\mu}}{\delta v^{\nu}},
\label{eq:ghost}
\end{equation}
where $v^\nu$ characterizes the vector field along which we study the
Lie derivative generating the coordinate transformations,
\begin{align}
 \frac{\delta \metric_{\alpha \beta}}{\delta v^{\nu}} = \frac{\delta}{\delta v^{\nu}} \mcL_{v} \metric_{\alpha \beta} = 2 \frac{\delta}{\delta v^{\nu}} D_{(\alpha} v_{\beta)} \text{.}
\end{align}
The corresponding variation of the gauge-fixing condition yields
\begin{align}
 \delta F^{\mu} = {2 \left( \bar{\metric}^{\mu \alpha} \bar{D}^{\beta} - \frac{(1+\beta)}{d} \bar{\metric}^{\alpha \beta} \bar{D}^{\mu} \right) D_{(\alpha} \delta v_{\beta)} \text{.}}
\end{align}
Let us decompose the vector $\delta v_{\beta}$ into a transversal part
$\delta v^{\mrT}_{\beta}$ and a longitudinal part $D_{\beta} \delta
\chi$. For the following argument, it suffices to study the limit of
the quantum metric approaching the background metric
$\metric_{\mu\nu} \to \bar{\metric}_{\alpha\beta}$, which
diagrammatically corresponds to studying the inverse ghost propagator
ignoring higher vertices
\begin{eqnarray}
 \delta F^{\mu} &=& (\delta^{\mu}_{\nu} \bar{D}^{2} \! + \! \bar{R}^{\mu}_{\nu}) \delta v^{\mrT \nu} \\
 &&+ \frac{1}{2} \big( ({d-1} \! - \! \beta) \bar{D}^{\mu} \bar{D}_{\nu} \! + \! 4 \bar{R}^{\mu}_{\nu} \big) \bar{D}^{\nu} \delta \chi  
+ \mathcal{O}(\metric -\bar\metric)\text{.}\nonumber
\end{eqnarray}
In this form it is obvious that the longitudinal direction
$\bar{D}^{\nu} \delta \chi$ is not affected by the gauge fixing for
$\beta = {d-1}$ to zeroth order in the curvature.  In other words, the
gauge fixing is not complete for this singular case $\beta_\text{sing}={d-1}$. This
singularity is correspondingly reflected by the ghost propagator. The
Faddeev-Popov operator in \Eqref{eq:ghost} reads
\begin{equation}
\mcM^{\mu}_{\point \nu} = 2 \bar{\metric}^{\mu\beta}  \bar{D}^\alpha D_{(\alpha} \metric_{\beta)\nu} - {2\frac{1 + \beta}{d}}
\bar{\metric}^{\alpha\beta} \bar{D}^{\mu} D_{\alpha} \metric_{\beta\nu}.
\end{equation}
Decomposing the ghost fields
$\bar{C}_{\mu}, C^{\nu}$ also into transversal $\bar{C}^{\mrT}_{\mu} ,
C^{\mrT \nu}$ and longitudinal parts $\bar{D}^{\mu} \bar{\eta} ,
\bar{D}^{\nu} \eta$ we find for the ghost Lagrangian,
\begin{eqnarray}
 \bar{C}_{\mu} \mcM^{\mu}_{\point \nu} C^{\nu} &=& \bar{C}^{\mrT}_{\mu} \left( \delta^{\mu}_{\nu} \bar{D}^{2} + \bar{R}^{\mu}_{\nu} \right) C^{\mrT \nu}
 \\
 &-&\!\bar{\eta} \left( \frac{{d-1} \! - \! \beta}{2} \bar{D}^{4} \! + \! \bar{R}^{\mu \nu} \bar{D}_{\mu} \bar{D}_{\nu} \right)\!\eta \nonumber + \mathcal{O}(\metric -\bar\metric)\text{.} \label{eq:invghost}
\end{eqnarray}
where we have performed partial integrations in order to arrive at a
convenient form and dropped covariant derivatives of the curvature.
This form of the inverse propagator of the ghosts makes it obvious
that a divergence of the form $\frac{1}{{d-1} - \beta}$ arises in the
longitudinal parts. This divergence at $\beta_{\text{sing}}={d-1}$ related
to an incomplete gauge fixing will be visible in all our results
below.

Let us now turn to the metric modes. As a technical tool, we
parametrize the fully dynamical metric $\metric_{\mu\nu}$ in terms of
a fiducial background metric $\bar\metric_{\mu\nu}$ and fluctuations
$\flucmet_{\mu\nu}$ about the background. Background independence is obtained
by keeping $\bar\metric_{\mu\nu}$ arbitrary and requiring that
physical quantities such as scattering amplitudes are independent of
$\bar\metric_{\mu\nu}$. Still, these requirements do not completely fix
the parametrization of the dynamical field
$\metric=\metric[\bar\metric;\flucmet]$. Several parametrizations have been
used in concrete calculations. The most commonly used parametrization
is the \textit{linear split} \cite{DeWitt:2003pm}
\begin{equation}
\metric_{\mu\nu}=\bar\metric_{\mu\nu} + \flucmet_{\mu\nu} \, .
\label{eq:linsplit}
\end{equation}
By contrast, the \text{exponential split} \cite{Kawai:1993fq,Kawai:1992np,
Kawai:1993mb,Kawai:1995ju,Aida:1994zc}
\begin{equation}
\metric_{\mu\nu}=\bar\metric_{\mu\rho} \big( e^\flucmet \big)^\rho{}_{\nu},
\label{eq:expsplit}
\end{equation}

is a parametrization that has been discussed more recently to a
greater extent \cite{Nink:2014yya,Demmel:2015zfa,Percacci:2015wwa,Labus:2015ska}.
In both cases, $\flucmet$ is
considered to be a symmetric matrix field (with indices raised and
lowered by the background metric). If a path integral of quantum
gravity is now defined by some suitable measure $\mathcal{D} \flucmet$, it is
natural to expect that the space of dynamical metrics $\metric$ is
sampled differently by the two parametrizations, implying different
predictions at least for off-shell quantities -- unless the variable
change from \eqref{eq:linsplit} to \eqref{eq:expsplit} is taken care
of by suitable (ultralocal) Jacobians. While a parametrization (and
gauge-condition) independent construction of the path integral has
been formulated in a geometric setting \cite{Fradkin:1983nw,Vilkovisky:1984st,
DeWitt:2003pm,Burgess:1987zi,Kunstatter:1991kw}, its
usability is hampered by the problem of constructing the full
decomposition of $\flucmet$ in terms of fluctuations between physically
inequivalent configurations and fluctuations along the gauge
orbit. Geometric functional \RG{} flows have been conceptually developed
in \cite{Pawlowski:2003sk}, with first results for asymptotic safety
obtained in \cite{Donkin:2012ud}, and recently to a leading-order
linear-geometric approximation in \cite{Demmel:2014hla}. The relation
between the geometric approach and the exponential parametrization was
discussed in \cite{Demmel:2015zfa}.

In the present work, we take a more pragmatic viewpoint, and consider
the different parameterizations of Eqs.~\eqref{eq:linsplit} and
\eqref{eq:expsplit} as
two different approximations of an ideal parametrization. Since the
functional \RG{} actually requires the explicit form of
$\metric[\bar\metric;h]$ only to second order in $\flucmet$ (in the
single-metric approximation, see below), we mainly consider a
one-parameter class of parametrizations of the type
\begin{align}
 \metric_{\mu \nu} = \bar{\metric}_{\mu \nu} + \flucmet_{\mu \nu} + \frac{\tau}{2} \flucmet_{\mu \rho} \flucmet^{\rho}_{\nu} + \mcO(\flucmet^3) \text{.}
\end{align}
For $\tau=0$, we obtain the linear split, whereas $\tau=1$ is
\textit{exactly} related to the exponential split within our
truncation.  Incidentally, it is straightforward to write down the
most general, ultra-local parametrization to second order that does
not introduce a scale,
\begin{align}
 \metric_{\mu \nu} &= \bar{\metric}_{\mu \nu} + \flucmet_{\mu \nu} \notag \\
 &+ \frac{1}{2} \left( {\tau} \flucmet_{\mu \rho} \flucmet^{\rho}_{\nu} + \tau_2 h \flucmet_{\mu\nu}
 + \tau_3 \bar{\metric}_{\mu\nu} \flucmet_{\rho\sigma} \flucmet^{\rho\sigma} + \tau_4 \bar{\metric}_{\mu\nu} h^2 \right) \notag \\
 &+ \mcO(\flucmet^3) \text{.}
 \label{eq:mostgenparam}
\end{align}
Here, $h = \flucmet_\mu^\mu$ is the trace of the fluctuation. As
mentioned above, third and higher-order terms will not contribute to
our present study anyway. Instead of exploring the full parameter
dependence, we will highlight some interesting results
in this more general framework below.

The key ingredient for a quantum computation is the propagator of the
dynamical field. In our setting, its inverse is given by the second
functional derivative (Hessian) of the action \eqref{eq:EH} including the gauge
fixing \eqref{eq:Sgf} with respect to the fluctuating field $\flucmet$,

\begin{align}
&\begin{aligned}
& \frac{1}{\mcZ_{R}} \left. \Gamma_{\flucmet \flucmet \point[2] \alpha \beta}^{(2) \kappa \nu} \right|_{\flucmet = 0, C = 0}
\\
& \! = \frac{1}{16 \alpha} \big( 8 \alpha \delta^{\kappa \nu}_{\alpha \beta} - [ 8 \alpha - (1 + \beta)^{2} ] \bar{\metric}^{\kappa \nu} \bar{\metric}_{\alpha \beta} \big) (- \bar{D}^{2})
\\
&\,\, - \frac{1 - \alpha}{\alpha} \delta^{(\kappa}_{(\alpha} \bar{D}^{\nu)} \bar{D}_{\beta)}
\\
&\,\, + \frac{1 + \beta - 2 \alpha}{4 \alpha} \big( \bar{\metric}^{\kappa \nu} \bar{D}_{(\alpha} \bar{D}_{\beta)} + \bar{\metric}_{\alpha \beta} \bar{D}^{(\kappa} \bar{D}^{\nu)} \big)
\\
&\,\, + \frac{1}{4} \big( 2 (1 - \tau) \delta^{\kappa\nu}_{\alpha \beta} - \bar{\metric}^{\kappa \nu} \bar{\metric}_{\alpha \beta} \big) (\bar{R} - 2 \lambda_{k})
\\
 &\,\, - ( 1 - \tau ) \bar{R}^{(\kappa}_{(\alpha} \delta^{\nu)}_{\beta)} + \frac{1}{2} (\bar{R}^{\kappa \nu} \bar{\metric}_{\alpha \beta} + \bar{R}_{\alpha \beta} \bar{\metric}^{\kappa \nu}) - \bar{R}^{\kappa \hphantom{(}\point \nu}_{\point (\alpha \point \beta)} \text{,}
\label{eq:Hessian}
\end{aligned}
\end{align}
Here and in the following, we specialize to $d=4$, except if stated
otherwise.  A standard choice for the gauge parameters is harmonic
DeDonder gauge with $\alpha=1=\beta$ for which the second and third
lines simplify considerably. Simplifications also arise for the
exponential split $\tau = 1$; in particular, a dependence on the
cosmological constant $\lambda_k$ remains only in the trace mode $\sim
\bar\metric^{\kappa\nu}\bar\metric_{\alpha\beta}$.

A standard tool for dealing with the tensor structure of the
propagator is the York decomposition of the fluctuations
$\flucmet_{\mu \nu}$ into transverse traceless tensor modes, a
transverse vector mode and two scalar modes,
\begin{align}
 & \flucmet_{\mu \nu} = \flucmet^{\mrT}_{\mu \nu} \! + \! 2 \bar{D}_{(\mu} \xi^{\mrT}_{\nu)} \! + \! \left( 2 \bar{D}_{(\mu} \bar{D}_{\nu)} \! - \! \frac{1}{2} \bar{\metric}_{\mu \nu} \bar{D}^{2} \right) \sigma \! + \! \frac{1}{4} \bar{\metric}_{\mu \nu} \flucmet \text{,}\label{eq:York1}
\\
 & \bar{D}^{\mu} \flucmet^{\mrT}_{\mu \nu} = 0, \quad \bar{\metric}^{\mu \nu} \flucmet^{\mrT}_{\mu \nu} = 0, \quad \bar{D}^{\mu} \xi^{\mrT}_{\mu} = 0 \text{.}\label{eq:York2}
\end{align}
It is convenient to split $\Gamma^{(2)}$ into a pure kinetic part
$\mcP$ which has a nontrivial flat-space limit, and a
curvature-dependent remainder $\mcF = \mcO(\bar{R})$. This facilitates
an expansion of the propagator $(\Gamma^{(2)})^{-1} = (\mcP +
\mcF)^{-1} = \sum\limits_{n=0}^{\infty} (- \mcP^{-1} \mcF)^{n}
\mcP^{-1}$.

Let us first concentrate on the kinetic part $\mcP$:
\begin{align}
 &\mcP_{\flucmet^\mrT \point[2] \alpha \beta}^{\point[2] \mu \nu} = \frac{\mcZ_{R}}{2} \delta^{\mu \nu}_{\alpha \beta} \big( \Delta - 2 (1 - \tau) \lambda_{k} \big) \text{,}
\\
 &\mcP_{\xi^\mrT \point \alpha}^{\point[2] \mu} = \frac{\mcZ_{R}}{\alpha} \delta^{\mu}_{\alpha} \Delta \big( \Delta - 2 \alpha ( 1 - \tau ) \lambda_{k} \big) \text{,} 
\\
&\mcP_{(\sigma \flucmet)} \! = \!\! \mcZ_{R} \!\! \begin{pmatrix} 3 \frac{(3 - \alpha) \Delta - 4 \alpha ( 1 - \tau) \lambda_{k}}{4 \alpha} \Delta^{\! 2} & \!\!\!\! \frac{3}{8 \alpha} (\beta \! - \! \alpha) \Delta^{\! 2} \vspace{0.05cm} \\ \frac{3}{8 \alpha} (\beta \! - \! \alpha) \Delta^{\! 2} & \!\!\!\! \frac{(\beta^{2} - 3 \alpha) \Delta + 4 \alpha ( 1 + \tau ) \lambda_{k}}{16 \alpha} \end{pmatrix}\! \text{,}
\label{eq:Psigmah}
\end{align}
where $\Delta = - \bar{D}^{2}$.  In this form it is straightforward to
calculate the propagator $(\mcP)^{-1}$. In particular, the transverse
traceless mode $\flucmet^\mrT$ does not exhibit any dependence on the
gauge parameters. As discussed in the introduction, a-priori criteria
suggest the Landau-gauge limit $\alpha\to0$ as a preferred choice for
the gauge fixing, as it strictly implements the gauge-fixing
condition. It thus should also be a fixed point of the \RG{} flow
\cite{Ellwanger:1995qf,Litim:1998qi}. Whereas the choice of $\alpha$ and
$\beta$, in principle, are independent, there can arise a subtle
interplay with certain regularization strategies as will be
highlighted in the following.

By taking the limit $\alpha \to 0$ while keeping $\beta$ finite,
we make the gauge fixing explicit, especially we find for the
gauge-dependent modes
\begin{align}
&  \mcP_{\xi^\mrT \point \alpha}^{-1 \mu} \! \to \alpha \frac{1}{\mcZ_{R} \Delta^{2}} \delta^{\mu}_{{\alpha}} \text{,}
\label{eq:scalar_propagator_alpha_to_zero1}\\
& \mcP_{(\sigma \flucmet)}^{-1} \! \to \! \frac{- \frac{1}{3 \mcZ_{R}} \Delta^{-2}}{\frac{(3 - \beta)^{2}}{4} \Delta \! - \! ( 3 \! - \! \beta^{2} \! + \! (3 \! + \! \beta^{2}) \tau) \lambda_{k}} \! \begin{pmatrix} \! \beta^{2} & \!\! - 6 \beta \Delta \\ \! - 6 \beta \Delta & \!\! 36 \Delta^{2} \end{pmatrix} \! \text{.}
\label{eq:scalar_propagator_alpha_to_zero2}
\end{align}
The transverse mode $\xi^{\mrT}_{\mu}$ decouples linearly with
$\alpha\to0$ and hence is pure gauge in the present setting.  Whereas
finite parts seem to remain in the $(\sigma\flucmet)$ subspace, we
observe that the matrix $\mcP_{(\sigma \flucmet)}^{-1}$ in
\eqref{eq:scalar_propagator_alpha_to_zero2} becomes degenerate in
this limit (e.g., the determinant of the matrix in
\Eqref{eq:scalar_propagator_alpha_to_zero2} is zero).  Effectively,
only one scalar mode remains in the propagator. The nature of this
scalar mode is a function of the second gauge parameter: taking the
limit $\beta \to \infty$, the remaining scalar mode can be identified
with $\sigma$, while the limit $\beta \to 0$ leaves us with a pure
$\flucmet$ mode.  

Whereas the transverse modes in
\Eqref{eq:scalar_propagator_alpha_to_zero1} decouple smoothly in the
limit $\alpha\to 0$, the decoupling of the scalar mode in
\Eqref{eq:scalar_propagator_alpha_to_zero2} is somewhat hidden in the
degeneracy of the scalar sector with the corresponding eigenmode
depending on $\beta$. This can lead to a subtle interplay with
regularization techniques for loop diagrams as can be seen on rather
general grounds by the following argument. Structurally, the
propagator in the $(\sigma\flucmet)$ sector has the following form in
the limit $\alpha\to0$ and for small but finite $\beta$, cf. \Eqref{eq:scalar_propagator_alpha_to_zero2}
\begin{align}
 \big( \mcP_{(\sigma \flucmet)} \big)^{-1} \to \begin{pmatrix} \mcO (\beta^{2}) & \mcO(\beta) \\ \mcO(\beta) & \mcO(1) \end{pmatrix} \! \text{.}
\end{align}
Regularizations of traces over loops built from this propagator are
typically adjusted to the spectrum of the involved operators. Let us
formally write this as
\begin{equation}
\Tr\,\Big[ \mathcal{L}_\mcR \,  \mcP^{-1} (\dots) \Big] 
\label{eq:sketchtrace}
\end{equation}
where $\mathcal{L}_\mcR$ denotes a regularizing operator and the
ellipsis stands for further vertices and propagators. Now, it is often
useful to regularize all fluctuation operators at the same scale,
e.g., the spectrum of all $\Delta$'s should be cut off at one and the same
scale $k^2$. Therefore, the regularizing operator $\mathcal{L}_\mcR$
inherits its tensor structure from the Hessian $\Gamma^{(2)}$ of
\Eqref{eq:Hessian}. In the $(\sigma\flucmet)$ sector, the regularizing
operator can hence acquire the same dependence on the gauge-parameters
as in \Eqref{eq:Psigmah},
\begin{align}
 \mathcal{L}_{\mcR,(\sigma\flucmet)} \to \frac{1}{\alpha} \begin{pmatrix} \mcO(1) & \mcO(\beta) \\ \mcO(\beta) & \mcO(\beta^{2}) \end{pmatrix} \! \text{,}
\end{align}
for $\alpha\to0$ and small $\beta$. The complete scalar contribution
to traces of the type \eqref{eq:sketchtrace} would then be of the
parametric form,
\begin{align}
 \Tr \left[ \mathcal{L}_\mcR \,   \mcP^{-1} (\dots) \right]_{(\sigma \flucmet)} \to \frac{1}{\alpha} \mcO(\beta^{2}) \text{.}\label{eq:alphadiv}
\end{align}
For finite $\beta$, such regularized traces can thus be afflicted with
divergencies in the Landau-gauge limit $\alpha\to0$. If this happens,
we still have the option to choose suitable values of $\beta$. In
fact, \Eqref{eq:alphadiv} suggest that still a whole one-parameter
family of gauges exists in the Landau-gauge limit, if we set
$\beta=\gamma\cdot \sqrt{\alpha}$, with arbitrary real but finite
gauge parameter $\gamma$ distinguishing different gauges.

We emphasize that this is a rather qualitative analysis. Since the
limit of products is not necessarily equal to the product of limits,
the trace over the matrix structure of the above operator products can
still eliminate this $1/\alpha$ divergence, such that any finite value
of $\beta$ remains admissible. 

In the following we observe that the appearance of the $1/\alpha$
divergence depends on the explicit choice of the regularization
procedure, as expected. Still, as this discussion shows, even if this
divergence occurs, it can perfectly well be dealt with by choosing
$\beta =\gamma \sqrt{\alpha}$ and still retaining a whole
one-parameter family of gauges in the Landau gauge limit.

\section{Gravitational RG flow}\label{sec:results}

For our study of generalized parametrization dependencies of
gravitational \RG{} flows, we use the functional \RG{} in terms of a
flow equation for the effective average action (Wetterich equation)
\cite{Wetterich:1992yh} amended by the background-field method
\cite{Reuter:1993kw,Reuter:1997gx,Freire:2000bq} and formulated for gravity \cite{Reuter:1996cp}
\begin{equation}
 \partial_t \Gamma_k[\metric,\bar\metric]=\frac{1}{2} \text{STr}\,\left[\partial_t \mcR_k \big(\Gamma_k^{(2)}+\mcR_k\big)^{-1} \right], \qquad \partial_t=k \frac{d}{dk}.
 \label{eq:floweq}
\end{equation}
Equation \eqref{eq:floweq} describes the flow of an action functional
$\Gamma_k$ as a function of an \RG{} scale $k$ that serves as a
regularization scale for the infrared fluctuations.  Here,
$\Gamma_k^{(2)}$ denotes the Hessian of the action with respect to the
fluctuation field $\metric$, at fixed background $\bar\metric$. The
details of the regularization are encoded in the choice of the
regulator $\mcR_k$. Suitable choices of $\mcR_k$ guarantee that
$\Gamma_k$ becomes identical to the full quantum effective action in
the limit $k\to 0$, and approaches the bare action for large scales
$k\to\Lambda_{\text{UV}}\to\infty$ (where $\Lambda_{\text{UV}}$ denotes a \UV{} cutoff). For
reviews in the present context, see
\cite{Reuter:1996ub,Pawlowski:2005xe,Gies:2006wv,Niedermaier:2006wt,Percacci:2007sz,Reuter:2012id,Nagy:2012ef}.

Whereas exact solutions of the flow equation so far have only been
found for simple models, approximate nonperturbative flows can be
constructed with the help of systematic expansion schemes. In the case
of gravity, a useful scheme is given by expanding $\Gamma_k$ in powers
of curvature invariants. The technical difficulties then lie in the
construction of the inverse of the regularized Hessian
$\big(\Gamma_k^{(2)}+\mcR_k\big)^{-1}$, corresponding to the
regularized propagator, and performing the corresponding traces (the
supertrace $\text{STr}$ includes a minus sign for Grassmann degrees of
freedom, i.e., Faddeev-Popov ghosts).

A conceptual difficulty lies in the fact that
$\Gamma_k[\metric,\bar\metric]$ should be computed on a subspace of
action functionals that satisfy the constraints imposed by
diffeomorphism invariance and background independence.  In general,
this requires to work with $\metric$ and $\bar\metric$ independently
during large parts of the computation
\cite{Manrique:2009uh,Manrique:2010am,Bridle:2013sra,Dietz:2015owa}. Such
bi-metric approaches can, for instance, be organized in the form of a
vertex expansion on a flat space as put forward recently in
\cite{Christiansen:2012rx,Christiansen:2014raa,Christiansen:2015rva},
or via a level expansion as developed in \cite{Becker:2014qya}, see
\cite{Codello:2013fpa,Becker:2014jua,Becker:2014pea} for further
bi-metric results.  For the present study of parametrization
dependencies, we confine ourselves to a single-metric approximation,
defined by identifying $\metric$ with $\bar\metric$ on both sides of
the flow equation, after the Hessian has been analytically
determined. In the following, we therefore do no longer have to
distinguish between the background field and the fluctuation field as
far as the presentation is concerned, and hence drop the bar notation
for simplicity.

Spanning the action in terms of the Einstein-Hilbert truncation
\eqref{eq:EH} and neglecting the flow of the gauge-fixing and ghost
sector \cite{Eichhorn:2009ah,Groh:2010ta,Eichhorn:2010tb}, we use the
\textit{universal \RG{} machine}
\cite{Benedetti:2010nr,Groh:2011vn,Groh:2011dw} as our computational
strategy. The key idea is to subdivide the Hessian $\Gamma_k^{(2)}$
into a kinetic part and curvature parts with a subsequent expansion in
the curvature. This is complicated by terms containing uncontracted
covariant derivatives in $\Gamma_k^{(2)}$ which could invalidate the
counting scheme. Within the present truncation, this problem is solved
with the aid of the York decomposition \eqref{eq:York1}. This helps
both to set up the curvature expansion as well as to invert the
kinetic terms in the corresponding subspaces of $\mathrm{TT}$,
$\mathrm{T}$ and scalar modes.  From a technical point of view,
  we use the package xAct \cite{xActwebpage,2007CoPhC.177..640M,
    2008CoPhC.179..586M,2008CoPhC.179..597M,Brizuela:2008ra,
    2014CoPhC.185.1719N} to handle the extensive tensor calculus.

Schematically, the flow equation for the Einstein-Hilbert truncation
can then be written as
\begin{equation}
\partial_t \Gamma_k=
{\int \text{d}^4x \sqrt{-\metric}}
\left(\mcS^{\text{TT}}+\mcS^{\text{T}}+\mcS^{\sigma
  h}+\mcS^{\text{gh}}+\mcS^{\text{Jac}}\right),
\label{eq:decompflow}
\end{equation}
where the first three terms denote the contributions from the graviton
fluctuations as parametrized by the York decomposition
\eqref{eq:York1}. The fourth term $\mcS^{\text{gh}}$ arises from the
Faddeev-Popov ghost fluctuations, cf. \Eqref{eq:invghost}. The last
term $\mcS^{\text{Jac}}$ comes from the use of transverse decompositions
of the metric \eqref{eq:York1} and the ghost fields
\eqref{eq:invghost}. The corresponding functional integral measure
over the new degrees of freedom involves Jacobians which -- upon
analogous regularization -- contribute to the flow of the effective
average action. 

At this point, we actually have a choice that serves as another source
of parametrization dependencies studied in this work: one option is to
formulate the regularized path integral in terms of the decomposed
fields as introduced above. In that case, the Jacobians are nontrivial
and their contribution $\mcS^{\text{Jac}}$ is listed in
\Eqref{eq:DefSJac}. Alternatively, we can reintroduce canonically
normalized fields by means of a nonlocal field redefinition
\cite{Dou:1997fg,Lauscher:2001ya},
\begin{eqnarray}
\sqrt{\Delta-\text{Ric}}\,\, \xi^\mu &\to& \xi^\mu, \\
\sqrt{\Delta^2+ \frac{4}{3}D_\mu R^{\mu\nu} D_\nu}\,\, \sigma & \to& \sigma, \\
\sqrt{\Delta}\,\, \eta & \to& \eta,\label{eq:fieldredef}
\end{eqnarray}
and analogously for the longitudinal anti-ghost field $\bar\eta$. (Here, we have used
$(\text{Ric}\,\xi)^\mu=R^{\mu\nu}\xi_\nu$.) This field redefinition goes 
along with another set of Jacobians contributing to the measure of the
rescaled fields. As shown in \cite{Lauscher:2001ya}, the Jacobians for
the original York decomposition and the Jacobians from the field
redefinition \eqref{eq:fieldredef} cancel at least on maximally
symmetric backgrounds. The latter choice of backgrounds is sufficient
for identifying the flows in the Einstein-Hilbert
truncation. Therefore, if we set up the flow in terms of the redefined
fields \eqref{eq:fieldredef}, the last term in \Eqref{eq:decompflow}
vanishes, $\mcS^{\text{Jac}}_{\text{fr}}=0$.

For an exact solution of the flow, it would not matter whether or not
a field redefinition of the type \eqref{eq:fieldredef} is
performed. Corresponding changes in the full propagators would be
compensated for by the (dis-)appearance of the Jacobians. For the present
case of a truncated nonperturbative flow, a dependence on the precise
choice will, however, remain, which is another example for a
parametrization dependence. This dependence also arises from the
details of the regularization. The universal \RG{} machine suggests to
construct a regulator $\mcR_k$ such that the Laplacians $\Delta$
appearing in the kinetic parts are replaced by
\begin{equation}
\Delta \to \Delta+R_k(\Delta), \label{eq:URGreg}
\end{equation}
where $R_k(x)$ is a (scalar) regulator function that provides a finite
mass-like regularization for the long-range modes, e.g., $R_k(x)\to
k^2$, for $x\ll k^2$, but leaves the \UV{} modes unaffected, $R_k(x)\to0$
for $x\gg k^2$. Since the field redefinition \eqref{eq:fieldredef} is
nonlocal, it also affects the kinetic terms and thus takes influence
on the precise manner of how modes are regularized via
\Eqref{eq:URGreg}. In other words, the dependence of our final results
on using or not using the field redefinition \eqref{eq:fieldredef} is
an indirect probe of the regularization-scheme dependence and thus of
the generalized parametrization dependence we are most interested in
here.

In this work, we focus on the \RG{} flow of the effective average action
parametrized by the operators of the Einstein-Hilbert truncation
\eqref{eq:EH}. For this, we introduce the dimensionless versions of
the gravitational coupling and the cosmological constant,
\begin{equation}
g:=\frac{k^2}{16\pi \mathcal{Z}_R}\equiv k^2 G, \quad \lambda= \frac{\Lambda}{k^2},
\end{equation}
and determine the corresponding \RG{} $\beta$ functions for $g$ and
$\lambda$, by computing the $\mcS$ terms on the right hand side of the
flow \eqref{eq:decompflow} to order $R$ in the curvature. Many
higher-order computations have been performed by now
\cite{Lauscher:2001rz,Lauscher:2002sq,Codello:2006in,Codello:2007bd,Machado:2007ea,Codello:2008vh,Bonanno:2010bt,Falls:2013bv,Benedetti:2009rx,Benedetti:2009gn,Rechenberger:2012pm,Demmel:2015oqa},
essentially confirming and establishing the simple picture visible in
the Einstein-Hilbert truncation.

We are particularly interested in the existence of fixed points
$g_\ast$ and $\lambda_\ast$ of the $\beta$ functions, defined by
\begin{equation}
\partial_t g=\dot g \equiv \beta_g(g_\ast,\lambda_\ast)=0, \quad 
\partial_t \lambda =\dot \lambda \equiv \beta_\lambda(g_\ast,\lambda_\ast)=0.
\end{equation}
In addition to the Gau\ss{}ian fixed point $g_\ast=0=\lambda_\ast$, we
search for a non-Gau\ss{}ian interacting fixed point, the existence of
which is a prerequisite for the asymptotic-safety scenario. Physically
viable fixed points should have a positive value for the Newton
coupling and should be connectable by an \RG{} trajectory with the
classical regime, where the dimensionful couplings are approximately
constant, i.e., the dimensionless versions should scale as $g \sim
k^2$, $\lambda \sim 1/k^2$. The asymptotic-safety scenario also
requires that a possible non-Gau\ss{}ian fixed point has finitely many
\UV{} attractive directions. This is quantified by the number of positive
critical exponents $\theta_i$ which are defined as $(-1)$ times the
eigenvalues of the stability matrix $\partial
\beta_{(g,\lambda)}/\partial (g,\lambda)$.

Whereas the fixed-point values $g_\ast$ and $\lambda_\ast$ are \RG{}
scheme-dependent, the critical exponents $\theta_i$ are universal and
thus should be parametrization independent in an exact
calculation. Also, the product $g_\ast\lambda_\ast$ has been argued to
be physically observable in principle and thus should be universal
\cite{Lauscher:2001ya}.  Testing the parametrization dependence of the
  critical exponents $\theta_i$ and $g_\ast\lambda_\ast$ therefore
  provides us with a quantitative criterion for the reliability of
  approximative results.

\section{Generalized parametrization dependence}

With these prerequisites, we now explore the parametrization
dependencies of the following scenarios: we consider the linear
\eqref{eq:linsplit} and the exponential \eqref{eq:expsplit} split,
both with and without field redefinition \eqref{eq:fieldredef}, and
study the corresponding dependencies on the gauge parameters,
focusing on a strict implementation of the gauge-fixing condition
$\alpha\to0$ (Landau gauge). As suggested by the principle of minimum
sensitivity, we look for stationary points as a function of the
remaining parameter(s) where universal results become most insensitive
to these generalized parametrizations. For the following quantitative
studies, we exclusively use the piecewise linear regulator
\cite{Litim:2001up,Litim:2002cf}, $R_k(x)=(k^2-x^2) \theta(k^2-x^2)$,
for reasons of simplicity. Studies of regulator-scheme dependencies
which can also quantify parametrization dependencies have first been
performed, e.g., in \cite{Lauscher:2001ya,Codello:2008vh}.

\subsection{Linear split without field redefinition}
\begin{figure}
\includegraphics[width=0.48\textwidth]{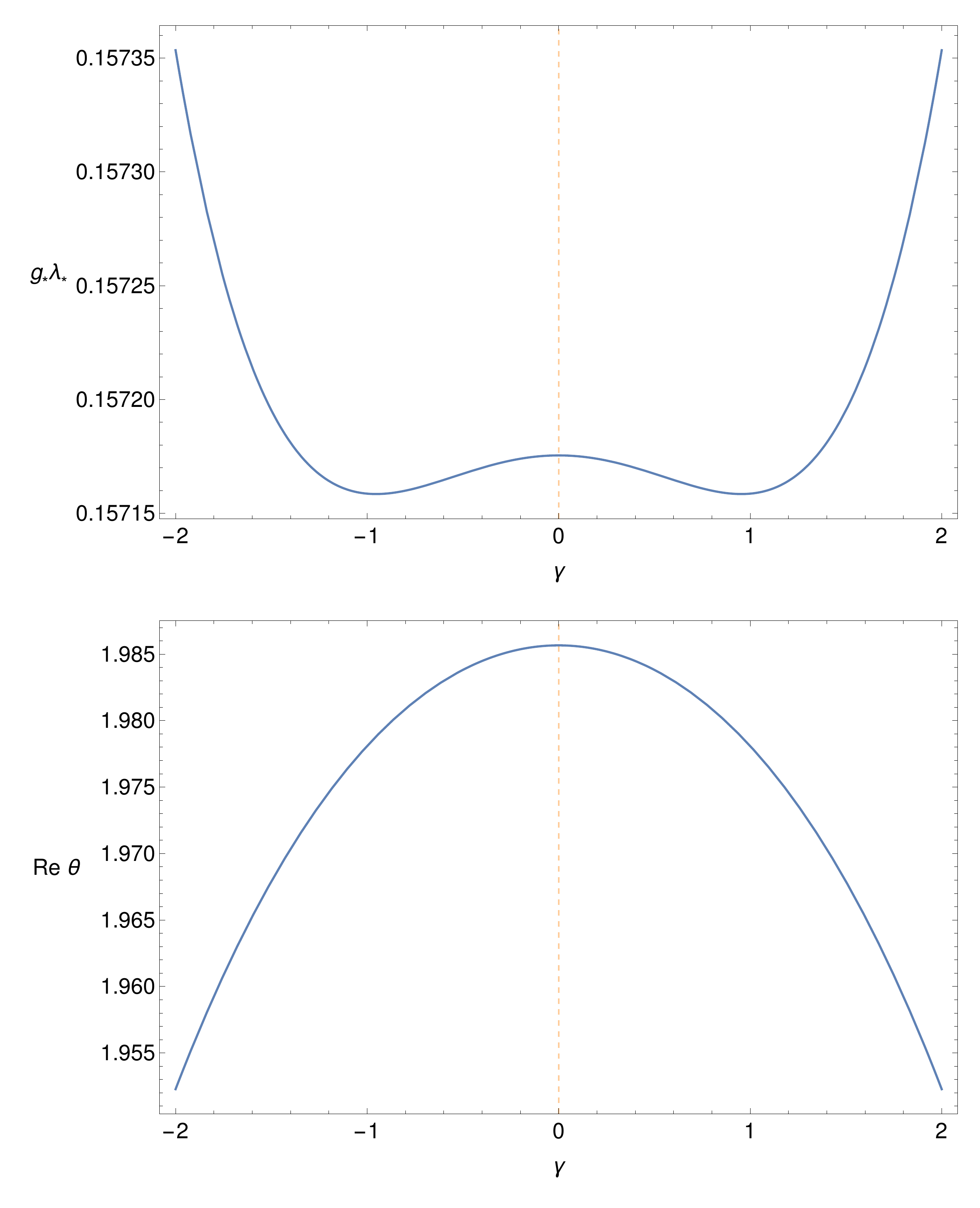}
\caption{Linear split without field redefinition: residual dependence
  of our estimates for the universal quantities on gauge parameter
  $\gamma$ in the limit $\alpha\to0$. We find a common stationary
  point at $\gamma=0$ and a remarkably small variation of the results on
  the level of $0.1\%$ for $g_\ast\lambda_\ast$ and $1.6\%$ for
  $\text{Re}\theta$ in the range $\gamma\in[-2,2]$.}
\label{fig:redef_phasediag}
\end{figure}

Let us start with the case of the linear split \eqref{eq:linsplit}
without field redefinition \eqref{eq:fieldredef}. Here, the degeneracy
in the sector of scalar modes interferes with the regularization
scheme, as illustrated in \Eqref{eq:alphadiv}. Hence, in the
Landau-gauge limit $\alpha\to0$, we choose $\beta=\gamma\,
\sqrt{\alpha}$, which removes any artificial divergence, but keeps
$\gamma$ as a real parameter that allows for a quantification of
remaining parametrization/gauge dependence. We indeed find a
non-Gau\ss{}ian fixed point $g_\ast,\lambda_\ast$ for wide range of
values of $\gamma$. The critical exponents form a complex conjugate
pair. The estimates for the universal quantities $g_\ast\lambda_\ast$
and the real part of the $\theta$'s (being the measure for the \RG{}
relevance of perturbations about the fixed point) are depicted in
Fig.~\ref{fig:redef_phasediag}.

We observe a common point of minimum sensitivity at $\gamma=0$. In a
rather wide range of gauge parameter values $\gamma\in [-2,2]$, our
estimates for $g_\ast\lambda_\ast$ and $\text{Re} \theta$ vary only
very mildy on the level of $0.1\%$ and $1.6\%$. Given the limitations
of the present simple approximation, this is a surprising degree of
gauge independence lending further support to the asymptotic-safety
scenario. The extremizing values at $\gamma=0$ are near the results of
\cite{Benedetti:2010nr,Codello:2007bd,Machado:2007ea} where the same
gauge choice ($\alpha=\beta=0$) was used. The main difference can be
traced back to the fact that our inclusion of the (dimensionful)
wave-function renormalization in the gauge fixing term \eqref{eq:Sgf}
renders the gauge parameter $\alpha$ dimensionless as is
conventional. If we ignored the resulting dimensional scaling, our
extremizing result would be exactly that of \cite{Benedetti:2010nr}
and in close agreement with \cite{Codello:2007bd,Machado:2007ea} with
slight differences arising from the regularization scheme. We
summarize a selection of our quantitative results in
Table~\ref{table:results}.

\begin{table}

\begin{tabular}{l|c|c|c|c}
 parametrization & $g_\ast$ & $\lambda_\ast$ & $g_\ast \lambda_\ast$ & $\theta$ \\ \hline\hline
 nfr $\tau=\alpha=\gamma=0$ & 0.879 & 0.179 & 0.157 & 1.986 $\pm$ \textbf{i} 3.064 \\ \hline
 nfr $\tau=0$, $\alpha=\beta=1$ & 0.718 & 0.165 & 0.119 & 1.802 $\pm$ \textbf{i} 2.352 \\ \hline
 fr  $\tau=\alpha=0$, $\beta=1$ & 0.893 & 0.164 & 0.147 & 2.034 $\pm$ \textbf{i} 2.691 \\ \hline
 fr  $\tau=0$, $\alpha=\beta=1$ & 0.701 & 0.172 & 0.120 & 1.689 $\pm$ \textbf{i} 2.486 \\ \hline
 fr  $\tau=\alpha=0$, $\beta=\infty$ & 0.983 & 0.151 & 0.148 & 2.245 $\pm$ \textbf{i} 2.794 \\ \hline
 fr $\tau=1$, $\beta=\infty$ & 3.120 & 0.331 & 1.033 & 4, 2.148 \\ \hline
 fr $\tau=1.22$, $\alpha=0$, $\beta=\infty$ & 3.873 & 0.389 & 1.508 & 3.957, 1.898
\end{tabular}  
\caption{Non-Gau\ss{}ian fixed-point properties for several
  parametrizations, characterized by the gauge parameters
  $\alpha,\beta$ or $\gamma$, as well as by the choice of the
  parametrization split parameter $\tau$ with $\tau=0$ corresponding
  to the linear split \eqref{eq:linsplit} and $\tau=1$, being the
  exponential split \eqref{eq:expsplit}. Whether or not a field
  redefinition \eqref{eq:fieldredef} is performed is labeled by
  ``fr'' or ``nfr'', respectively.}\label{table:results}
 
\end{table}

\subsection{Exponential split without field redefinition}
\begin{figure}
\includegraphics[width=0.48\textwidth]{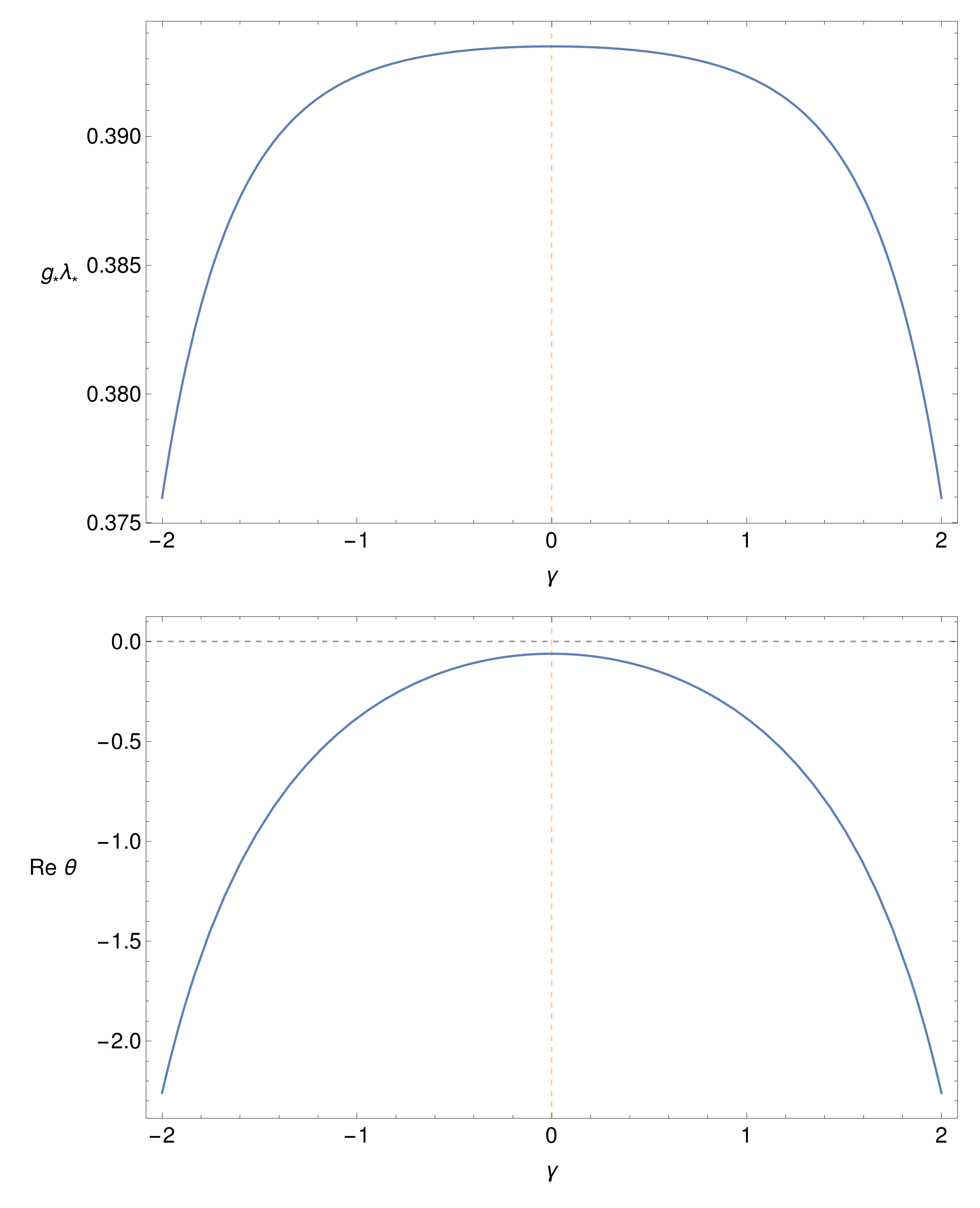}
\caption{Exponential split without field redefinition: residual
  dependence of our estimates for the universal quantities on gauge
  parameter $\gamma$ in the limit $\alpha\to0$. A common stationary
  point is again present at $\gamma=0$, but the estimates for the
  universal quantities exhibit a substantial variation in the range
  $\gamma\in[-2,2]$: $g_\ast\lambda_\ast$ varies by $\sim 5\%$ and
  $\text{Re}\theta$ even by more than a factor of {40}. The latter is a
  clear signal for the insufficiency of the parametrization.}
\label{fig:nofieldredef_expsplit_glambdatheta_alpha0_gammadep}
\end{figure}

As a somewhat contrary example, let us now study the case of the
exponential split \eqref{eq:expsplit} also without field redefinition
\eqref{eq:fieldredef}. Again, we find a non-Gau\ss{}ian fixed
point. The corresponding estimates for the universal quantities at
this fixed point in the Landau gauge limit $\alpha=0$ are displayed in
Fig.~\ref{fig:nofieldredef_expsplit_glambdatheta_alpha0_gammadep}. At
first glance, the results seem similar to the previous ones with a
stationary point at $\gamma=0$. However, the product
$g_\ast\lambda_\ast$ shows a larger variation on the order of 5\% and
the critical exponent even varies by a factor of more than {40} in the
range $\gamma\in[-2,2]$. We interpret the strong dependence on the
gauge parameter $\gamma$ as a clear signature that these estimates
based on the exponential split without field redefinition should not
be trusted.

In fact, the real part of the critical exponents, $\text{Re}\theta$,
have even changed sign compared to the previous case implying that the
non-Gau\ss{}ian fixed point has turned \UV{} repulsive. Similar
observations have been made in \cite{Nink:2014yya} for the harmonic
Feynman-type gauge $\alpha=1=\beta$ and an additional strong
dependence on the regulator profile function $R_k(x)$ has been
found. We have verified that our results agree with those of
\cite{Nink:2014yya} for the corresponding gauge choice. In summary,
this parametrization serves as an example that non-perturbative
estimates can depend strongly on the details of the parametrization
(even for seemingly reasonable parametrizations) and the results can
be misleading. The good news is that a study of the parametrization
dependence can -- and in this case does -- reveal the insufficiency of
the parametrization through its strong dependence on a gauge
parameter.

\subsection{Linear split with field redefinition}

For the remainder, we consider parametrizations of the fluctuation
field which include field redefinitions \eqref{eq:fieldredef}. The
canonical normalization achieved by these field redefinitions has not
merely aesthetical reasons. An important aspect is that the nonlocal
field redefinition helps to regularize the modes in a more symmetric
fashion: the kinetic parts of the propagators then become linear in
the Laplacian which are all equivalently treated by the regulator
\eqref{eq:URGreg}. A practical consequence is that the interplay of
the degeneracy in the scalar sector no longer interferes with the
regularization, i.e., the gauge parameter $\beta$ can now be chosen
independently of $\alpha$. Concentrating again on the Landau-gauge
limit $\alpha\to0$, we observe for generic split parameter $\tau$ that
$\beta=0$ no longer is an extremal point.

\begin{figure}
\includegraphics[width=0.48\textwidth]{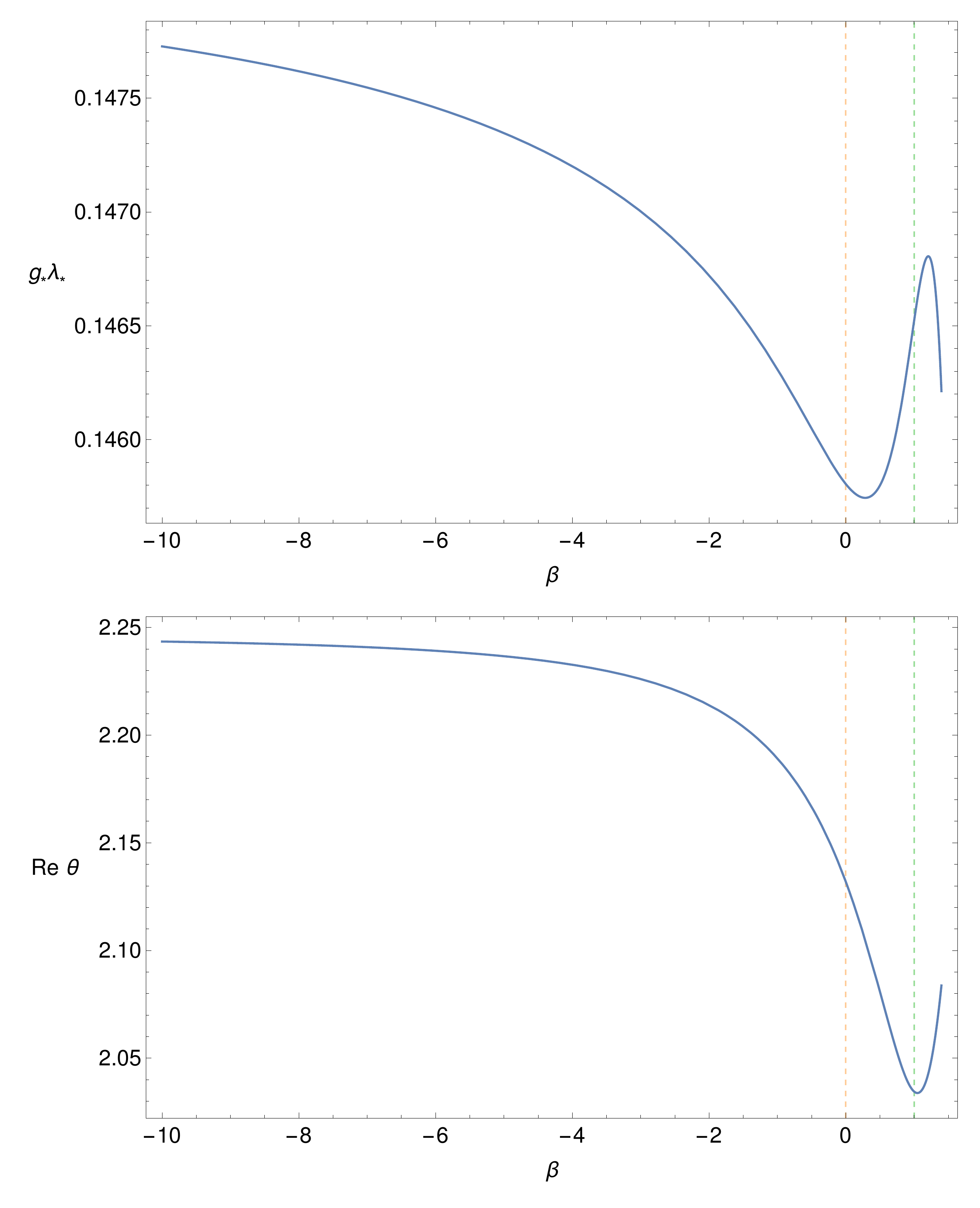}
\caption{Linear split with field redefinition: residual dependence of
  our estimates for the universal quantities on the gauge parameter
  $\beta$ in the limit $\alpha\to0$. A common stationary point is
  approached for $|\beta|\to\infty$. Near the harmonic gauge $\beta=1$
  (green dashed vertical line), both quantities have an extremum. For
  the whole range of $\beta$ values, the estimates for the universal
  quantities exhibit rather small variations of $1\%$ for
  $g_\ast\lambda_\ast$ and $10\%$ for the more sensitive critical
  exponent $\text{Re}\theta$.}
\label{fig:fieldredef_linsplit_glambdatheta_alpha0_betadep}
\end{figure}

Our estimates for the universal quantities for the case of the linear
split \eqref{eq:linsplit} with field redefinition
\eqref{eq:fieldredef} and $\alpha\to0$ are plotted in
Fig.~\ref{fig:fieldredef_linsplit_glambdatheta_alpha0_betadep}. In
order to stay away from the singularity at $\beta=3$,
cf. \Eqref{eq:invghost}, we consider values for $\beta<3$ down to
$\beta\to-\infty$. As is obvious, e.g., from
\Eqref{eq:scalar_propagator_alpha_to_zero2}, the dependence of the
propagator of the scalar modes and thus on $\beta$ is such that the
limits of large positive or negative $\beta\to\pm\infty$ yield
identical results. Also the longitudinal ghost mode decouples in the
limit $\beta\to\pm\infty$ such that the whole flow in the large
$|\beta|$-limit is independent of the sign of $\beta$. 

A non-Gau\ss{}ian fixed point exists, and a common extremum of
$g_\ast\lambda_\ast$ and $\text{Re}\, \theta$ occurs for
$\beta\to-\infty$. Near $\beta=1$ marking the harmonic gauge
condition, both quantities are also close to an extremum (which does
not occur at exactly the same $\beta$ value for both quantities). All
fixed-point quantities for this case are listed in
Tab.~\ref{table:results} (``fr {$\tau=\alpha=0$}, $\beta=1$''). These
values agree with the results of \cite{Donkin:2012ud}.
They are remarkably close, e.g., to those for the linear split
without field redefinition. The situation is similar for the other
extremum $|\beta|\to\infty$ (``fr {$\tau=\alpha=0$, $\beta=\infty$}'' in
Tab.~\ref{table:results}). For the whole infinite $\beta$ range
studied for this parametrization, $g_\ast\lambda_\ast$ varies on the
level of 1\%. The more sensitive critical exponent $\text{Re}\,\theta$
varies by 10\% which is still surprisingly small given the simplicity
of the approximation. Let us emphasize again that varying $\beta$ from
infinity to zero corresponds to a complete exchange of the scalar
modes from $\sigma$ (longitudinal vector component) to $h$ (conformal
mode) and hence to a rather different parametrization of the
fluctuating degrees of freedom.

\subsection{Exponential split with field redefinition}

\begin{figure}
\includegraphics[width=0.48\textwidth]{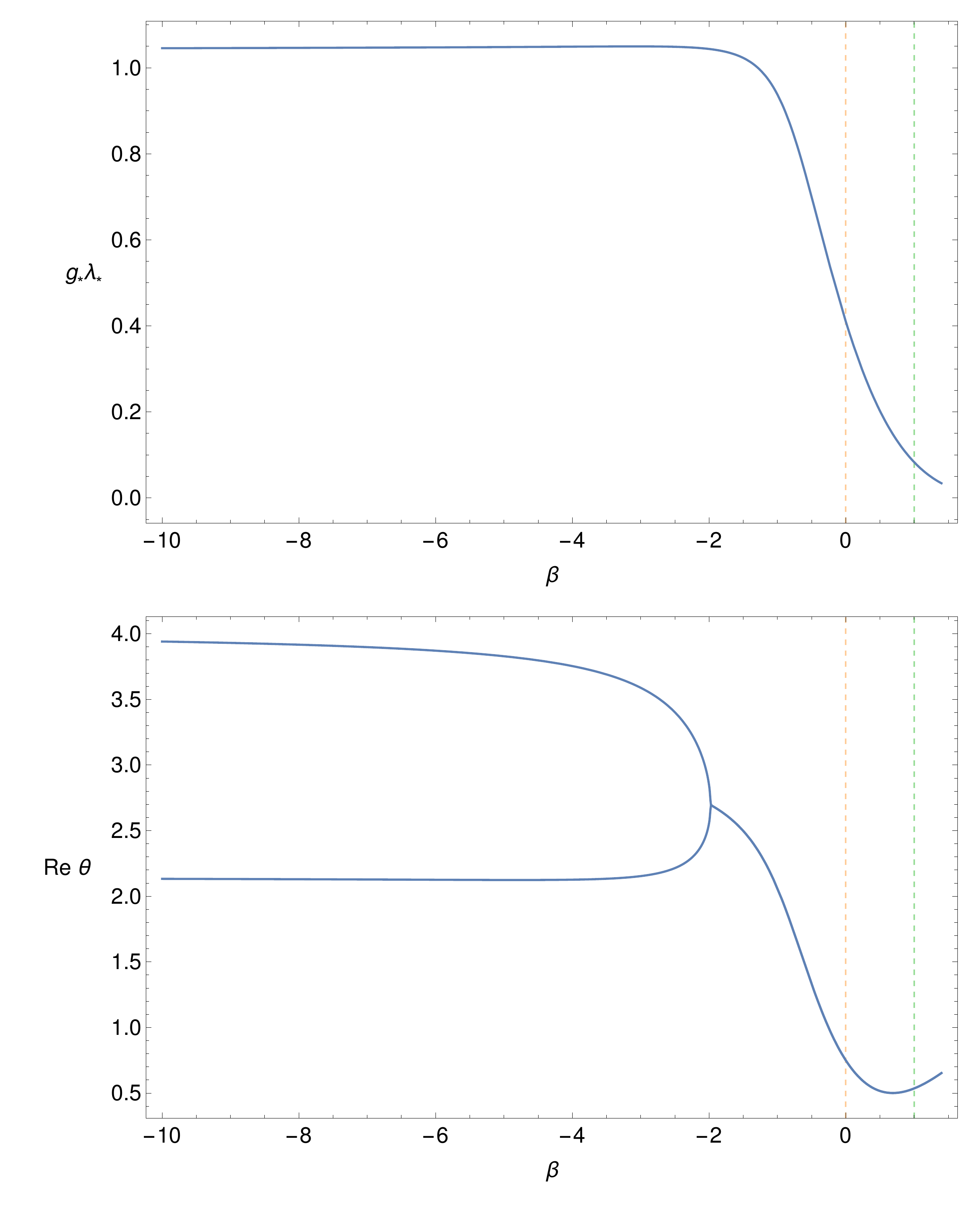}
\caption{Exponential split with field redefinition: residual
  dependence of our estimates for the universal quantities on the
  gauge parameter $\beta$ in the limit $\alpha\to0$. A common
  stationary point is approached for $|\beta|\to\infty$, whereas no
  common minimum-sensitivity point is found near the harmonic gauge
  $\beta=1$ or $\beta=0$ (dashed vertical lines). Below $\beta\lesssim
  -2$, the critical exponents become real with the non-Gau\ss{}ian
  fixed point remaining \UV{} attractive. For $|\beta|\to\infty$, the
  results become independent of the gauge parameter $\alpha$.}
\label{fig:fieldredef_expsplit_glambdatheta_alpha0_betadep}
\end{figure}

Finally, we consider the exponential split \eqref{eq:linsplit},
$\tau=1$, with field redefinition \eqref{eq:fieldredef}. Having
performed the latter has a strong influence on the stability of the
estimates of the universal quantities at the non-Gau\ss{}ian fixed
point, as is visible in
Fig.~\ref{fig:fieldredef_expsplit_glambdatheta_alpha0_betadep}. Contrary
to the linear split, we do not find a common extremum near small
values of $\beta$: neither $\beta=0$ nor the harmonic gauge $\beta=1$
seem special, but, e.g., the product $g_\ast\lambda_\ast$ undergoes a
rapid variation in this regime.

Rather, a common extremal point is found in the limit
$\beta\to\infty$. In fact, $g_\ast\lambda_\ast$ becomes insensitive to
the precise value of $\beta$ for $\beta\lesssim -2$ (with a local
maximum near $\beta\simeq -3$,
and an asymptotic value of $g_\ast\lambda_\ast\simeq 1.033$ for
$\beta\to\infty$. This estimate for $g_\ast\lambda_\ast$ is
significantly larger than for the other parametrizations. The
deviation may thus be interpreted as the possible level of accuracy
that can be achieved in this simple Einstein-Hilbert truncation. 

As an interesting feature, the critical exponents become real for
$\beta\lesssim -2$,
and approach the asymptotic values $\theta=\{4,2.148\}$ for
$\beta\to\infty$. The leading exponent $\theta=4$ reflects the
power-counting dimension of the cosmological term. This is a
straightforward consequence of the fact that the $\lambda$ dependence
in this parametrization $\tau=1$, $\beta\to\infty$ disappears from the
propagators of the contributing modes. The leading nontrivial exponent
$\theta=2.148$ hence is associated with the scaling of the Newton
constant near the fixed point, which is remarkably close to minus the
power-counting dimension of the Newton coupling. The latter is a
standard result for non-Gau\ss{}ian fixed points which are described
by a quadratic fixed-point equation
\cite{Gies:2003dp,Gehring:2015vja}. The small difference to the value
$\theta=2$ arises from the \RG{}-improvement introduced by the
anomalous dimension in the threshold functions (``$\eta$-terms'' as
discussed in the Appendix). Neglecting these terms, the estimate of
the leading critical exponents in dimension $d$ is $d$ and $d-2$, as
first discussed in \cite{Percacci:2015wwa}. Also our other
quantitative results for the fixed-point properties are in agreement
with those of \cite{Percacci:2015wwa} within the same approximation.

The significance of the results within this parametrization is further
underlined by the observation that the results in the limit
$\beta\to\infty$ become completely independent of the gauge parameter
$\alpha$. In other words, the choice of the transverse traceless mode
and the $\sigma$ mode ($\beta\to\infty$) as a parametrization of the
physical fluctuations removes any further gauge dependence. 

The present parametrization has also some relation to
\cite{Falls:2015qga,Falls:2015cta}, where in addition to the
exponential split the parametrization was further refined to remove
the gauge-parameter dependence completely on the semi-classical
level. More specifically, the parametrization of the fluctuations was
chosen so that only fluctuations contribute that also have an on-shell
meaning. In essence, this removes any contribution from the scalar
modes to the \UV{} running. At the semi-classical level
\cite{Falls:2015qga}, the nontrivial critical exponent is $2$ as in
\cite{Percacci:2015wwa} and increases upon inclusion of \RG{} improvement
as in the present work. The increase determined in
\cite{Falls:2015cta} is larger than in the present parametrization and
yields $\theta\simeq 3$ which is remarkably close to results from
simulations based on Regge calculus
\cite{Hamber:1999nu,Hamber:2015jja}.

The present parametrization with $|\beta|\to\infty$ is also loosely
related to unimodular gravity, as the conformal mode is effectively
removed from the fluctuation spectrum. Still, differences to
unimodular gravity remain in the gauge-fixing and ghost sector as
unimodular gravity is only invariant under transversal
diffeomorphisms. It is nevertheless interesting to observe that
corresponding \FRG{} calculations yield critical exponents of
comparable size \cite{Eichhorn:2013xr,Eichhorn:2015bna}.

In fact, the present parametrization allows for a closed form solution
of the \RG{} flow as will be presented in Sect.~\ref{sec:analytical}.

\subsection{Landau vs. Feynman gauge}

\begin{figure}
\includegraphics[width=0.48\textwidth]{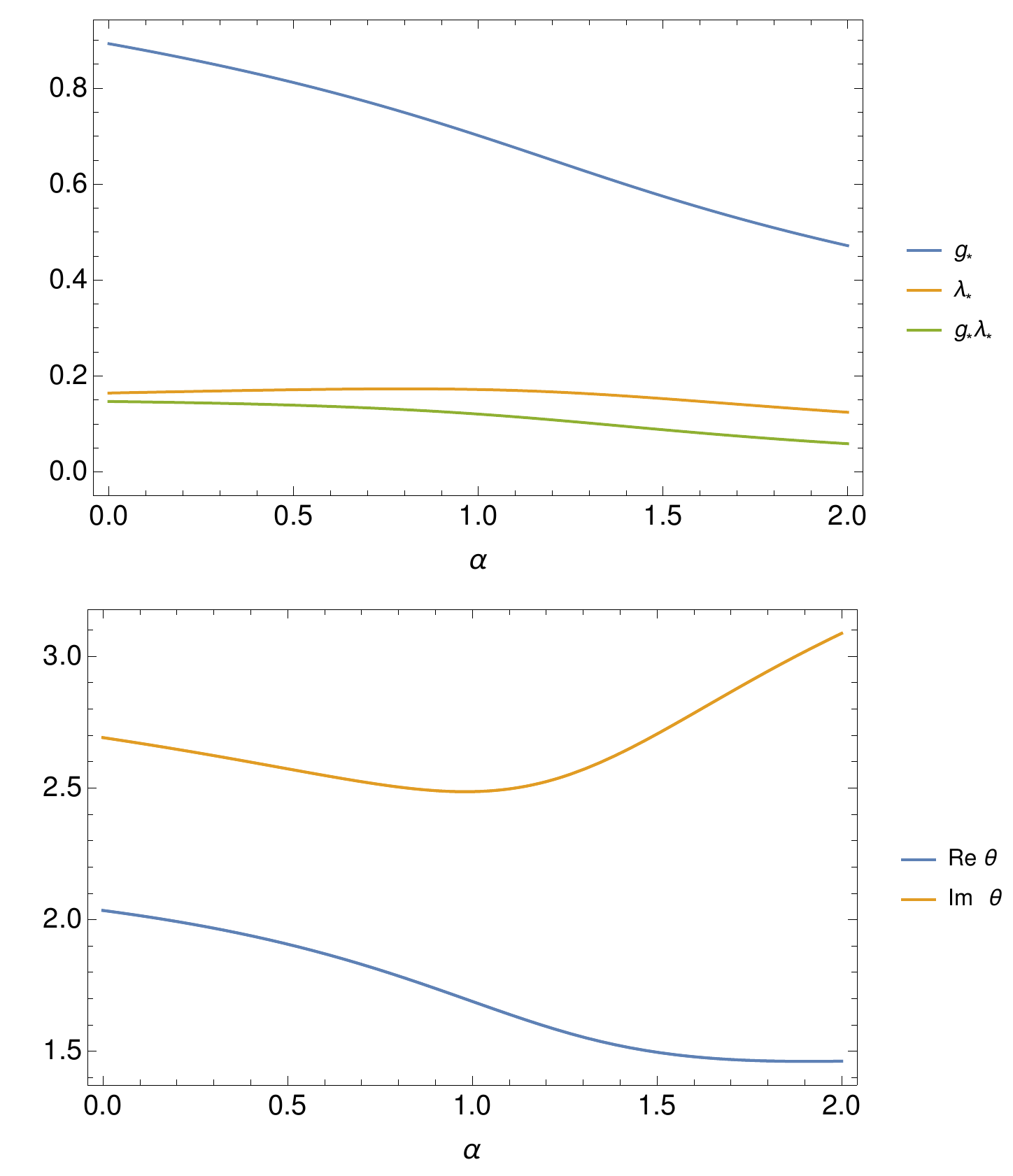}
\caption{Linear split with field redefinition: dependence of
  estimates for the fixed-point values (upper panel) and the critical
  exponents (lower panel) on the gauge parameter $\alpha$ and harmonic
  gauge condition $\beta=1$. No qualitative and only minor
  quantitative differences are found for the Feynman gauge $\alpha=1$
  in comparison to the Landau gauge $\alpha=0$.}
\label{fig:fieldredef_linsplit_alphadep_beta1}
\end{figure}

Many of the pioneering computations in quantum gravity have been and
still are performed within the harmonic gauge $\beta=1$ and with
$\alpha=1$ corresponding to Feynman gauge. This is because this choice
leads to a number of technical simplifications such as the direct
diagonalization of the scalar modes as is visible from the
off-diagonal terms in \Eqref{eq:Psigmah}. Concentrating on the linear
split with field redefinition, we study the $\alpha$ dependence for
the harmonic gauge $\beta=1$ in the vicinity of the Landau and Feynman
gauges. The results for the non-Gau\ss{}ian fixed point values are
shown in the upper panel of
Fig.~\ref{fig:fieldredef_linsplit_alphadep_beta1}. In essence, the
fixed-point values show only a mild variation during the transition
from the Landau gauge $\alpha=0$ to the Feynman gauge $\alpha=1$. In
particular, the decrease of $g_\ast$ is slightly compensated for by a
mild increase of $\lambda_\ast$. Effectively, the observed variation is 
only on a level which is quantitatively similar to other
parametrization dependencies, cf. Table~\ref{table:results}. 

A similar conclusion holds for the more sensitive critical
exponents. Real and imaginary parts of the complex pair are shown in
the lower panel of
Fig.~\ref{fig:fieldredef_linsplit_alphadep_beta1}. Starting from
larger values of $\alpha$, it is interesting to observe that the
imaginary part $\text{Im}\,\theta$ decreases with decreasing
$\alpha$. This may be taken as an indication for a tendency towards
purely real exponents; however, at about $\alpha=1$ this tendency is
inverted and the exponents remain a complex pair in between Feynman
gauge and Landau gauge within the present estimate.

In summary, we observe no substantial difference between the results
in Feynman gauge $\alpha=1$ and those of Landau gauge $\alpha=0$ in
any of the quantities of interest for the linear split and with field
redefinition. Our results show an even milder dependence on the gauge
parameter in comparison to the pioneering study of
Ref.~\cite{Lauscher:2001ya}, where the regulator was chosen such as to
explicitly lift the degeneracy in the sector of scalar modes in the
limit $\alpha\to0$. The present parametrization hence shows a
remarkable degree of robustness against deformations away from the
a-priori preferable Landau gauge. Hence, we conclude that the use of
Feynman gauge is a legitimate option to reduce the complexity of
computations.

\subsection{Generalized parametrizations}
\label{sec:generalized}

\begin{figure}
\includegraphics[width=0.48\textwidth]{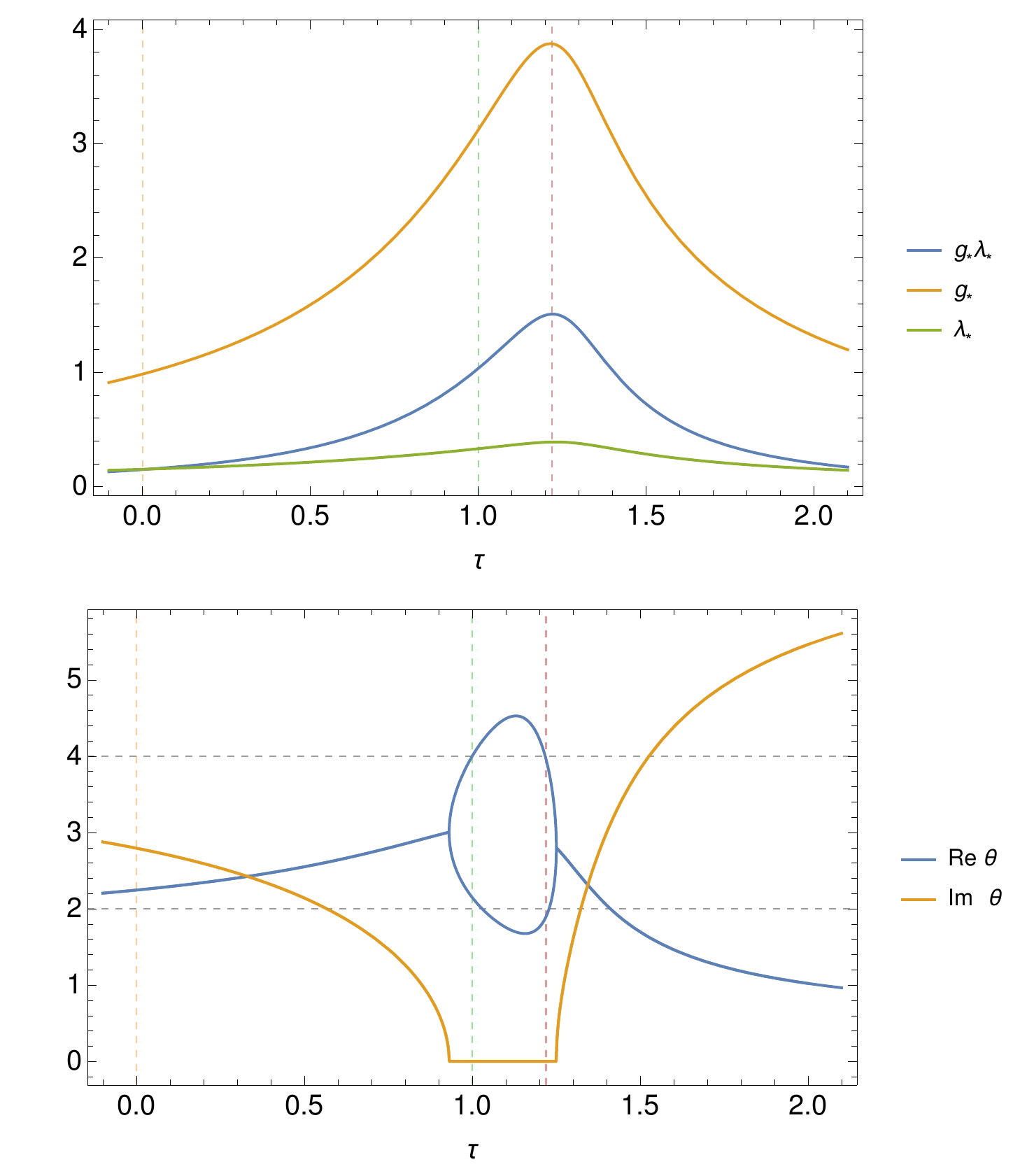}
\caption{Parametrization dependence of fixed-point values (upper
  panel) and critical exponents (lower panel) as a function of the
  split parameter $\tau$ for the Landau gauge $\alpha=0$ and
  $|\beta|\to \infty$. The fixed-point values exhibit extrema near
  $\tau\simeq 1.22$, for the product of
  fixed-point values, this occurs at 
  $\tau = 1 + \frac{\sqrt{3}}{24} \left( \frac{278}{\pi} \right)^{1/4}$
  (red dashed vertical line). In this regime, the
  critical exponents are real and close to their values for the
  exponential split $\tau=1$ (green dashed vertical line).}
\label{fig:fieldredef_glambdatheta_alpha0_betainf_splitdep}
\end{figure}

Having focused so far mainly on the gauge-parameter dependencies for
fixed values of the split parameter $\tau$, we now explore
the one-parameter family of parametrizations for general $\tau$. For
this, we use the Landau gauge $\alpha=0$ and take the limit
$|\beta|\to\infty$, where the fixed-point estimates of all
parametrizations used so far showed a large degree of stability.
Figure~\ref{fig:fieldredef_glambdatheta_alpha0_betainf_splitdep}
exhibits the results for the non-Gau\ss{}ian fixed-point values (upper
panel) and the corresponding critical exponents (lower panel).

A comparison of the results for $\tau=0$ and $\tau=1$ reveals the
differences already discussed above: an increase of the fixed-point
values and the occurrence of real critical exponents for the
exponential split $\tau=1$. From the perspective of the principle of
minimum sensitivity, it is interesting to observe that the fixed-point
values develop extrema near $\tau\simeq 1.22$. The product
$g_\ast\lambda_\ast$ is maximal for $\tau = 1 + \frac{\sqrt{3}}{24}
\left( \frac{278}{\pi} \right)^{1/4}$. Also for this parametrization,
the critical exponents of the fixed point are real and still close to
the values for the exponential split,
cf. Table~\ref{table:results}. For even larger values of $\tau$, the
critical exponents form complex pairs again.

To summarize, in the full three-parameter space defined by $\tau$,
$\beta$ and $\alpha\geq0$, we find a local extremum, i.e., a point of
minimum sensitivity, at $\alpha=0$, $\beta\to\infty$ and $\tau$ near
the exponential split value $\tau=1$. From this a-posteriori
perspective, our results suggest that the exponential split (with
field redefinition) in the limit where the scalar sector is
represented by the $\sigma$ mode may be viewed as a ``best estimate''
for the \UV{} behavior of quantum Einstein gravity. Of course, due to
the limitations imposed by the simplicity of our truncation, this
conclusion should be taken with reservations. The resulting \RG{} flow
for $\tau=1$ is in fact remarkably simple and will be discussed next.

\subsection{Analytical solution for the phase diagram}
\label{sec:analytical}

Let us now analyze more explicitly the results for the \RG{} flow for the
exponential split with field redefinition in the Landau gauge and in
the limit $|\beta|\to\infty$. Several simplifications arise in this
case. The exponential split removes any dependence of the transverse
traceless and vector components of the propagator on the cosmological
constant. The remaining dependence on $\lambda$ in the conformal mode
is finally removed by the limit $|\beta|\to \infty$. As a consequence,
the cosmological constant does not couple into the flows of the Newton
coupling nor into any other higher-order coupling. Still, the
cosmological constant is driven by graviton fluctuations. As
emphasized above, any remaining gauge dependence on the gauge
parameter $\alpha$ drops out of the flow equations. For the \RG{} flow of
Newton coupling and cosmological constant, we find the simple set of
equations:
\begin{align}
 \dot g\equiv\beta_g &= 2g -
 \frac{135g^2}{72\pi-5g} \label{eq:simpleflowg}\\ 
\dot\lambda\equiv\beta_\lambda
 &= \left(-2 - \frac{135g}{72\pi-5g} \right)\lambda - g \left(
 \frac{43}{4\pi} - \frac{810}{72\pi-5g} \right)
 \, \label{eq:simpleflowl}.
\end{align}
In addition to the Gau\ss{}ian fixed point, these flow equations support a fixed point at
\begin{equation}
g_\ast= \frac{144 \pi}{145}, \quad \lambda_\ast=\frac{48}{145}, \quad g_\ast\lambda_\ast= \frac{6912 \pi}{21025},
\label{eq:simpleflowFP}
\end{equation}
cf. Table~\ref{table:results}. Also the critical exponents $\theta_i$
being $(-1)$ times the eigenvalues of the stability matrix $\partial
\beta_{(g,\lambda)}/\partial (g,\lambda)$ can be determined
analytically,
\begin{equation}
\theta_0=4, \quad \theta_1=\frac{58}{27}.
\end{equation}
The fact that the largest critical exponent corresponds to the
power-counting canonical dimension of the cosmological term is a
straightforward consequence of the structure of the flow equations
within this parametrization: as we have $\dot g = (2 + \eta(g))g$ and
$\dot\lambda= (-2+\eta(g))\lambda + \mathcal{O}(g)$, the existence of
a non-Gau\ss{}ian fixed point requires $\eta(g_\ast)=-2$. As the
stability matrix is triangular, the eigenvalue associated with the
cosmological term must be $-4$ and thus $\theta_0=4$. Rather
generically, other parametrizations lead to a dependence of $\eta$
also on $\lambda$ and thus to a more involved stability matrix.

In the physically relevant domain of positive gravitational coupling
$g>0$, the fixed point $g_\ast$ separates a ``weak'' coupling phase
with $g<g_\ast$ from a ``strong'' coupling phase $g>g_\ast$. Only the
former allows for trajectories that can be interconnected with a
classical regime where the dimensionless $g$ and $\lambda$ scale
classically, i.e., $\dot g\simeq 2g$ and $\dot\lambda\simeq -2\lambda$
such that their dimensionful counterparts approach their observed
values. Trajectories in the strong-coupling phase run to larger values
of $g$ and terminate in a singularity of $\beta_g$ at
$g_\text{sing}=72\pi/5$ indicating the break-down of the truncation.

\begin{figure*}
\includegraphics[width=\textwidth]{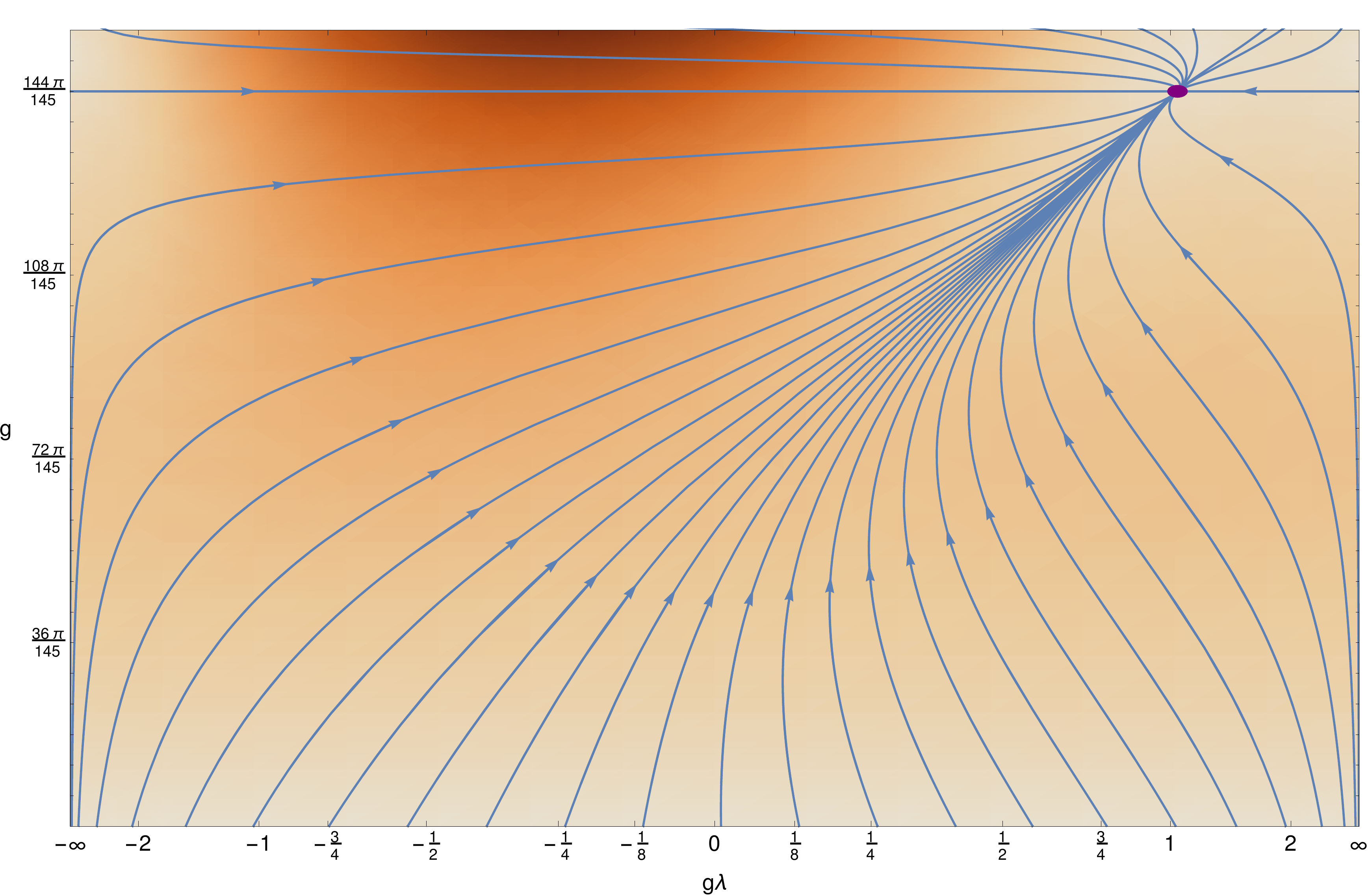}
\caption{Global phase diagram in the ($g,g\lambda$) plane for the
  exponential split with field redefinition and
  $|\beta|\to\infty$. Arrows point from \IR{} to \UV{} indicating the
  approach to the \UV{} fixed point at $g_\ast=144\pi/145$ and
  $\lambda_\ast=48/145$. The color indicates a measure for the flow velocity,
  $(\partial_t g)^2 + (\partial_t(g \lambda/\sqrt{1+g^2\lambda^2}))^2$.}
\label{fig:fieldredef_phasediag}
\end{figure*}

All trajectories in the weak coupling phase with $g<g_\ast$ run
towards the Gau\ss{}ian fixed point for $g$ and thus, also the flow of
$\lambda$ in the infrared is dominated by the Gau\ss{}ian fixed
point. This implies that all trajectories emanating from the
non-Gau\ss{}ian fixed point with $g\leq g_\ast$ can be continued to
arbitrarily low scales, i.e., are infrared complete. They can thus be
labeled by their deep infrared value of $g\lambda$ approaching a
constant, which may be identified with the product of Newton coupling
and cosmological constant as observed at present. A plot of the
resulting \RG{} flow in the plane
$(g,g\lambda)$ is shown in Fig.~\ref{fig:fieldredef_phasediag}. It
represents a global phase diagram of quantum gravity as obtained in
the present truncation/parametrization. We emphasize that no
singularities appear towards the \IR{} contrary to conventional
single-metric calculations based on the linear split.

The flows \eqref{eq:simpleflowg} and \eqref{eq:simpleflowl} can be
integrated analytically. Converting back to dimensionful
couplings, the flow of the running Newton coupling $G(k)$ satisfies the implicit equation,
\begin{equation}
G_N= \frac{G(k)}{\left( 1-\frac{145}{144\pi} k^2 G(k) \right)^{\frac{27}{29}}}, 
\end{equation}
where $G_N$ is the Newton coupling measured in the deep infrared
$k\to0$. Expanding the solution at low scales about the Newton
coupling yields
\begin{equation}
G(k)\simeq G_N \left( 1 - \frac{15}{16\pi} k^2 G_N  + \mathcal{O}\big( (k^2 G_N)^2 \big) \right)
\end{equation}
exhibiting the anti-screening property of gravity.

The flow of the dimensionful running cosmological constant
$\Lambda(k)$ can be given explicitly in terms of that of the running
Newton coupling,
\begin{align}
 \Lambda(k) &= \frac{162k^2}{25} - \frac{43 G(k) k^4}{16\pi} + \ell k^2 \left( 144\pi-145 G(k) k^2 \right)^\frac{25}{29} \notag \\
 &- \frac{144\pi}{3625 G(k)}\left( 87+25\ell \left( 144\pi-145 G(k) k^2 \right)^\frac{25}{29} \right) \, .
\end{align}
Here, $\ell = -\frac{29}{86400}\left( 2^{-13} 3^{-21}\pi^{-54} \right)^\frac{1}{29} (125\Lambda G_N+432\pi)$,
and $\Lambda$ is the value of the classical cosmological constant in the deep infrared $k\to 0$.
The low-scale expansion about $k=0$ yields
\begin{equation}
 \Lambda(k) \simeq \Lambda \left( 1 - \frac{15}{16 \pi} k^{2} G_N + \mcO \left( \frac{k^{4}}{\Lambda^{2}} \Lambda G_N , (k^2 G_N)^{2} \right) \right)
\end{equation}
Thus, $\Lambda(k)/G(k) = \Lambda/G_N + \mathcal O(k^4)$, implying a
comparatively slow running of the ratio towards the \UV{}. This explicit
solution of the \RG{} flow might be useful for an analysis of
``\RG{}-improved'' cosmologies along the lines of \cite{Bonanno:2001xi,Bonanno:2002zb,Hindmarsh:2011hx,Koch:2010nn,Babic:2004ev,Reuter:2005kb,Reuter:2012xf}.

\subsection{Generalized ultra-local parametrizations}
\label{sec:genparam2}

For the most general, ultra-local parametrization
\eqref{eq:mostgenparam}, it turns out that the flow equation in our
truncation does only depend on the linear combinations $T_1 :=
\tau/4+\tau_3$ and $T_2 := \tau_2/4+\tau_4$, leaving only two
independent split parameters. Instead of exploring the full high-dimensional
parameter space, we try to identify relevant points as inspired by our
preceding results. For instance for the choice $T_1=1/4$ and
$T_2=-1/8$, any dependence on $\alpha$ drops out, indicating an
enhanced insensitivity to the gauge choice. The resulting flow
equations are
\begin{align}
 \dot g &= 2g + \frac{135(\beta-3)g^2}{(5\beta-3)g-72(\beta-3)\pi} \, , \\
 \dot \lambda &= -2\lambda + \frac{g((-669+215\beta)g +36(\beta-3)\pi(4-15\lambda))}{4\pi((3-5\beta)g+72(\beta-3)\pi)} \, .
\end{align}
In the limit $|\beta|\to\infty$, these are identical to the
exponential split in the same limit. The non-Gau\ss{}ian fixed point
occurs at
\begin{align}
g_\ast= \frac{144 \pi (\beta-3)}{145\beta-411}, \quad \lambda_\ast=\frac{48(\beta-3)}{145\beta-411}, \notag \\
\quad g_\ast\lambda_\ast= 6912 \pi \left(\frac{\beta-3}{145\beta-411}\right)^2 \, .
\end{align}
Apart from the pathological choice $\beta_\text{sing}={3}$ (incomplete gauge fixing)
where this fixed point merges with the Gau\ss{}ian fixed point, no
further extremal point is observed except for the limit
$|\beta|\to\infty$. The critical exponents are
\begin{equation}
\theta_0=4, \quad \theta_1=\frac{58}{27} + \frac{16}{45(\beta-3)}.
\end{equation}
Also the exponents become minimally sensitive to the choice of $\beta$
for $|\beta|\to\infty$.  

As an oddity, we mention the particular case
$\beta=3/5$, where the flow equations acquire a pure one-loop form.
In this case, the second critical exponent is exactly $2$ as it must.
 
More importantly, the interdependence of gauge and parametrization
choices is also visible in the following fact: we observe that the
choice of the gauge parameter $|\beta|\to\infty$ removes any
dependence of our flow on the parameter $T_2$ independently of
the value of $\alpha$. In other words, this limit brings us back
exactly to the case which we discussed above in
Sect.~\ref{sec:generalized}, such that the seemingly much larger class
of parametrizations \eqref{eq:mostgenparam} collapses to a
one-parameter family.

\subsection{Arbitrary dimensions}
\label{sec:arbdim}

Finally, we discuss the stability of the \UV{} fixed-point scenario
and its parametrization dependence in arbitrary dimensions, focusing
on $d>2$ (for a discussion of $d=2$ in the present context, see
\cite{Nink:2014yya,Percacci:2015wwa,Falls:2015qga}). In fact, there are
some indications in the literature that the parametrization dependence
is pronounced in higher dimensions. Whereas standard calculations
based on the linear split generically find a \UV{} fixed point in any
dimension $d>2$ and gauge-fixing parameter $\alpha$, see
e.g. \cite{Litim:2003vp,Fischer:2006fz}, a recent refined choice of
the parametrization to remove gauge-parameter dependence on the
semi-classical level arrives at a different result
\cite{Falls:2015qga,Falls:2015cta}: the \UV{} fixed point can be
removed from the physical region if the number of physical gravity
degrees of freedom becomes too large. As the latter increases with the
dimensionality, there is a critical value $d_{\text{cr}}$ above which
asymptotically safe gravity does not exist. The resulting scenario is
in line with the picture of paramagnetic dominance
\cite{Nink:2012vd,Nink:2012kr}, which is also at work for the QED and
QCD $\beta$ functions: the dominant sign of the $\beta$ function
coefficient arises from the paramagnetic terms in the Hessian which
can be reversed if too many diamagnetically coupled degrees of freedom
contribute.

Our results extend straightforwardly to arbitrary
dimensions. Starting, for instance, with the most general
parametrization \eqref{eq:mostgenparam} in $d$ dimensions, the flows
of $g$ and $\lambda$ depend only on the linear combinations $T_1 =
\tau/d + \tau_3$ and $T_2 = \tau_2/d+\tau_4$. Comparable results as in
$d=4$ dimensions apply: in the limit of $|\beta|\to\infty$, also $T_2$
drops out such that a one-parameter family remains. In turn, a
complete independence of the gauge parameter $\alpha$ can be realized
with the parametrization specified by $T_1 = 1/d$ and $T_2 = -1/(2d)$.

\begin{figure}
\includegraphics[width=0.48\textwidth]{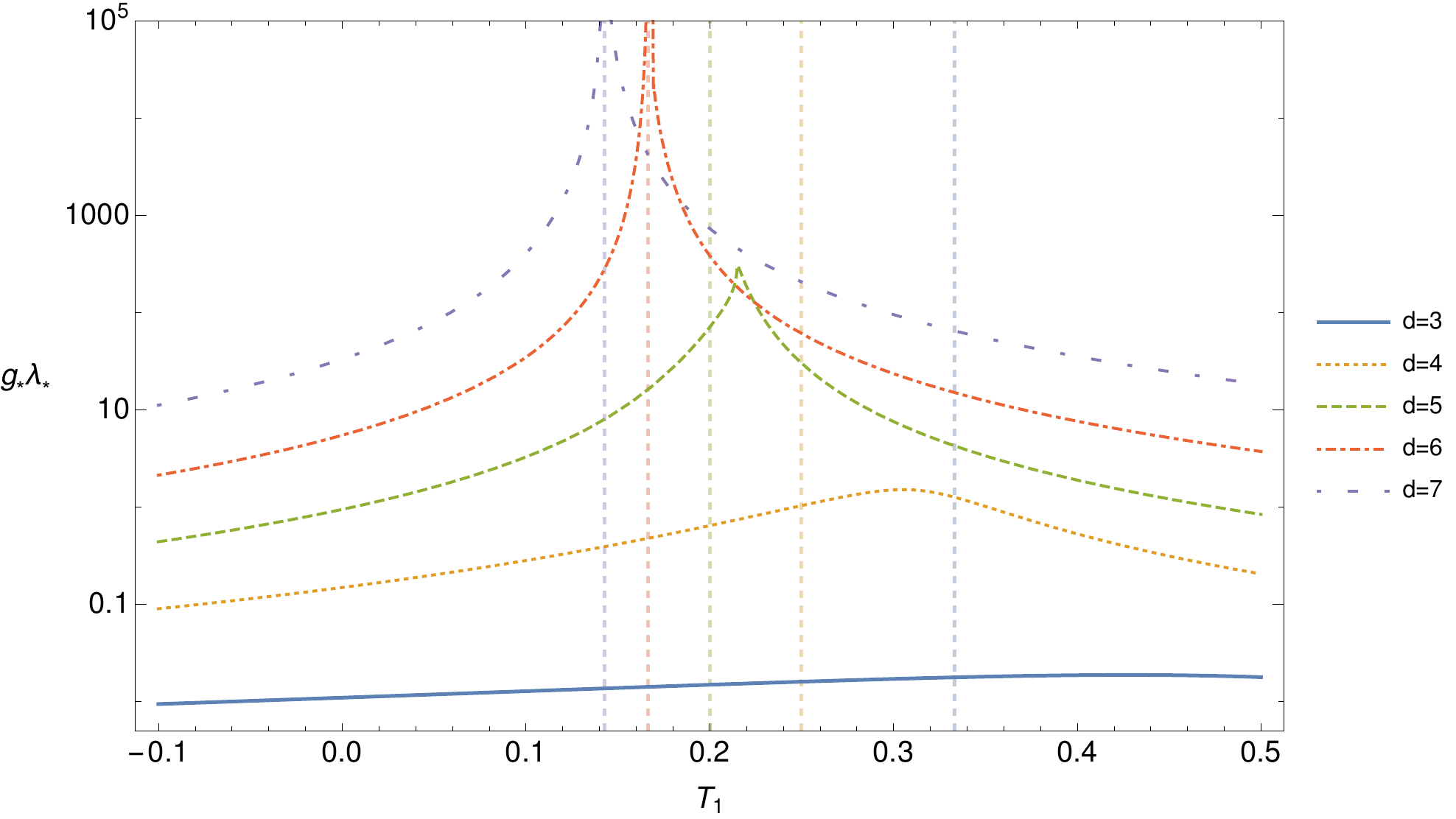}
\caption{Parametrization dependence of fixed-point value for
  $g_\ast\lambda_\ast$ as a function of the split parameter
  $T_1$ in the Landau gauge $\alpha=0$ and $|\beta|\to \infty$ for
  different dimensions $d=3,4,5,6,7$ (from bottom to top). Vertical
  lines mark the value of the parameter $T_1 = 1/d$ preferred by
  gauge-parameter $\alpha$ independence. For $d\geq
  d_{\text{cr}}\simeq 5.731$, the fixed-point product
  $g_\ast\lambda_\ast$ develops a singularity at $T_1=1/d$.}
\label{fig:fieldredef_dimdep}
\end{figure}

We illustrate the stability properties of the asymptotic-safety
scenario in arbitrary dimensions by choosing the Landau-gauge limit
$\alpha\to0$ as well as $|\beta|\to\infty$, keeping $T_1$ as a free
parameter. Then, we know a priori that $T_1=1/d$ would be a preferred
choice from the view point of gauge invariance; it would also
correspond to the exponential parametrization $\tau=1$,
$\tau_3=0$. Fig.~\ref{fig:fieldredef_dimdep} displays the fixed-point
values for $g_\ast\lambda_\ast$ as a function of $T_1$ for various
dimensions $d=3,\dots,7$. While $d=3$ exhibits a rather small
parametrization dependence, $d=4$ reproduces the earlier results of
Fig.~\ref{fig:fieldredef_glambdatheta_alpha0_betainf_splitdep} (upper
panel) now as a function of $T_1$ with an extremum not far above
$T_1=1/4$. By contrast, $g_\ast\lambda_\ast$ develops a kink for $d=5$
that turns into a singularity for $d=6$ and larger. For increasing
$d$, the kink approaches the preferred parametrization $T_1=1/d$
(vertical dashed lines in Fig.~\ref{fig:fieldredef_dimdep}). The
singularity in $g_\ast\lambda_\ast$ occurs for a critical dimension
$d_{\text{cr}}\simeq 5.731$.

This observation suggests the following interpretation: whereas we can
identify a \UV{} fixed point for any dimension as long as we choose $T_1$
sufficiently far away from $T_1=1/d$, we find a stable fixed-point
scenario only for $d=3$ and $d=4$ integer dimensions. Already for
$d=5$, the fixed-point product $g_\ast\lambda_\ast$ can change by two
orders of magnitude by varying the parametrization, which is at least
a signature for the insufficiency of the truncation. For $d\geq
d_{\text{cr}}\simeq 5.731$, $g_\ast\lambda_\ast$ can become unboundedly
large as a function of the parametrization, signaling the instability
of the fixed point.

If these features persist also beyond our truncation, they suggest
that the asymptotic safety scenario may not exist far beyond the
spacetime dimension $d=4$. Whereas this does not offer a dynamical
explanation of our spacetime dimension, it may serve to rule out the
mutual co-existence of extra dimensions and asymptotically safe
quantum gravity.

\section{Conclusions}

We have reexamined generalized parametrization dependencies of
non-perturbative computations in quantum gravity based on the
functional renormalization group. Whereas parametrically-ordered
expansion schemes such as perturbation theory for on-shell quantities
are free from such dependencies, off-shell quantities and
non-perturbative expansions rather generically exhibit dependencies
on, e.g., the choice of the regularization, the gauge fixing or the
field parametrization. In this work, we have dealt with these
dependencies in a pragmatic manner, analyzing the sensitivity and
stability of the \UV\ behavior of metric quantum gravity with respect to
variations of such generalized parametrizations. We have focused on
the question of the existence and the properties of a non-Gau\ss{}ian
UV fixed point, facilitating metric quantum gravity to be
asymptotically safe. We have also concentrated on a widely studied and
rather well-understood computing scheme, the Einstein-Hilbert
truncation in the single-metric formulation.

Our results show a remarkable stability in a variety of qualitative
aspects: for all parametrizations that exhibit rather large stationary
regimes in the space of all parameters, we have found a
non-Gau\ss{}ian \UV\ fixed point with two \RG\ relevant directions,
corresponding to the Newton coupling and the cosmological constant
being physical parameters. For most parametrizations, the universal
quantities show a remarkably mild (given the simplicity of the
approximation) variation and thus a high degree of stability. Our scan
of parametrization dependencies can also help identifying less robust
parametrizations, and thus help judging the physical relevance of
results.

Some features, however, appear to depend more strongly on the
parametrization or are even visible only in specific
parametrizations. Moreover, a nontrivial interplay between various
aspects of parametrizations, e.g., gauge choice vs. field
parametrization, can arise. With hindsight, the results obtained
within the exponential split with field redefinition in the limit
where the graviton degrees of freedom are spanned by the transverse
traceless and a scalar mode ($|\beta|\to\infty$) exhibit the highest
degree of comprehensiveness: complete independence of the gauge
parameter $\alpha$, fully analytical and integrable global flows with
a classical \IR\ limit in the physical parameter regime, real critical
exponents at the \UV\ fixed point, and the existence of an upper
critical dimension for the asymptotic safety scenario. The exploration
of higher-order truncations \cite{Ohta:2015efa} and the inclusion of
matter degrees of freedom
\cite{Percacci:2002ie,Eichhorn:2011pc,Dona:2013qba} in this
parametrization appears highly worthwhile,
c.f. \cite{Percacci:2015wwa,Labus:2015ska} for scalar matter.

In summary, our work exemplifies that a careful investigation of
parametrization dependencies facilitates both a test of the robustness
of nonperturbative quantum gravity computations as well as the
identification of a parametrizations which may be better adapted to
the physical mechanisms.

\section*{Acknowledgements}
We would like to thank Astrid Eichhorn, Kevin Falls, Jan Pawlowski,
Roberto Percacci, Gian-Paolo Vacca and Omar Zanusso for discussions.  This
work was supported by the DFG-Research Training Group ``Quantum- and
Gravitational Fields'' GRK 1523/2. HG acknowledges funding by the DFG under
grant no. Gi 328/7-1. BK acknowledges funding by the DFG
under grant no. Wi 777/11-1.

\appendix

\section{Flow equations}

In this section, we display the right hand side of the Wetterich
equation for general Regulators $R_k[\Delta]$ and in dimension
$d=4$. For simplicity, we introduce the anomalous dimension $\eta =
(\dot g-2g)/g$, and refer to terms linear in $\eta$ as
``$\eta$-terms''.  Let us start with the contribution from the
TT-mode:
\begin{align}
 \mathcal S^\text{TT} &= \frac{5}{2} Q_2 \left[ \frac{\dot R_k - \eta R_k}{\Delta + R_k - 2\lambda(1-\tau)} \right] \notag \\
 &-\frac{5}{12} R \left( Q_1 \left[ \frac{\dot R_k - \eta R_k}{\Delta + R_k - 2\lambda(1-\tau)} \right] \right. \\
 &\left. + (4-3\tau) Q_2 \left[ \frac{\dot R_k - \eta R_k}{(\Delta + R_k - 2\lambda(1-\tau))^2} \right] \right) \notag \, ,
\end{align}
where the $Q$ functionals are defined in terms of Mellin transforms
\cite{Reuter:1996cp}.  For the transverse vector, and without field
redefinition, let us define
\begin{align}
 \mathcal G_n^\text{1T} &= \left[ -(\dot R_k - \eta R_k) \left( 2\lambda (1-\tau) - \frac{1}{\alpha} (R_k+2\Delta) \right) \right. \notag \\
 &\qquad\left. -2(\dot\lambda+2\lambda)(1-\tau)R_k \right] \times \\
 &\qquad\left( (\Delta+R_k)\left(\frac{\Delta+R_k}{\alpha}+2\lambda(1-\tau)\right) \right)^{-n} \, . \notag
\end{align}
With that, we have
\begin{align}
 \mathcal S^\text{1T} &= \frac{3}{2} Q_2 \left[ \mathcal G_1^\text{1T} \right] + 
 R \left( \frac{1}{4} Q_1 \left[ \mathcal G_1^\text{1T} \right] \right.  \\ 
 &\left. + \frac{3}{2}(1-\alpha(1-\tau)) Q_3 \left[ \mathcal G_2^\text{1T} \right] - \frac{3}{4} \lambda (1-\tau) Q_2 \left[ \mathcal G_2^\text{1T} \right] \right) \, .\notag
\end{align}
On the other hand, the contribution with field redefinition reads, 
\begin{align}
 \mathcal S^\text{1T}_\text{fr} &= \frac{3}{2} Q_2 \left[ \frac{\dot R_k - \eta R_k}{\Delta+R_k-2\alpha\lambda(1-\tau)} \right] \notag \\
 &\!+ R \left( \frac{1}{8} Q_1 \left[ \frac{\dot R_k - \eta R_k}{\Delta+R_k-2\alpha\lambda(1-\tau)} \right] \right. \\
 &\!\left.+ \frac{3}{8}(1-2\alpha(1-\tau)) Q_2 \! \left[ \frac{\dot R_k - \eta R_k}{(\Delta+R_k-2\alpha\lambda(1-\tau))^2} \right] \right) \, . \notag
\end{align}
For the scalar contribution, we first define
\begin{align}
 \pi^\sigma &= -\frac{3}{4\alpha} \left( 4\alpha\lambda(1-\tau)(\Delta+R_k)^2 + (\alpha-3)(\Delta+R_k)^3 \right) \\
 \pi^\text{h} &= -\frac{1}{16\alpha} \left( -4\alpha\lambda(1+\tau) + (3\alpha-\beta^2)(\Delta+R_k) \right) \\
 \pi^\text{x} &= -\frac{3}{8\alpha}(\alpha-\beta)(\Delta+R_k)^2 \\
 \rho^\sigma &= -\frac{3}{4\alpha} \left( 4\alpha (\dot\lambda+(2-\eta)\lambda)(1-\tau)(2\Delta + R_k)R_k \right. \notag  \\
 &\qquad+8\alpha\lambda(1-\tau)(\Delta+R_k)\dot R_k \notag \\
 &\qquad+ 3(\alpha-3)(\Delta+R_k)^2 \dot R_k \notag \\
 &\left.\qquad- (\alpha-3)(3\Delta^2 + 3 \Delta R_k + R_k^2) \eta R_k \right) \\
 \rho^\text{h} &= -\frac{1}{16\alpha}(3\alpha-\beta^2)(\dot R_k - \eta R_k) \\
 \rho^\text{x} &= -\frac{3}{8\alpha}(\alpha-\beta)(2\dot R_k (\Delta+R_k) - \eta R_k (2\Delta+R_k)).
\end{align}
The contribution is
\begin{widetext}
\begin{eqnarray}
 S^{\sigma\text{h}} &=& \frac{1}{2} Q_2 \left[ \frac{\pi^\sigma\rho^\text{h} +
 \pi^\text{h} \rho^\sigma - 2 \pi^\text{x} \rho^\text{x}}{\pi^\sigma \pi^\text{h} - \left( \pi^\text{x} \right)^2} \right] 
 +R \left\{ \frac{1}{12} Q_1 \left[ \frac{\pi^\sigma\rho^\text{h} +
 \pi^\text{h} \rho^\sigma - 2 \pi^\text{x} \rho^\text{x}}{\pi^\sigma \pi^\text{h} - \left( \pi^\text{x} \right)^2} \right] \right. 
 -\frac{3}{4\alpha}(6-\alpha(4-3\tau)) Q_4 \left[ \frac{\rho^\text{h}}{\pi^\sigma \pi^\text{h} - \left( \pi^\text{x} \right)^2} \right] \notag \\
 &&+\lambda(1-\tau) Q_3 \left[ \frac{\rho^\text{h}}{\pi^\sigma \pi^\text{h} - \left( \pi^\text{x} \right)^2} \right] 
 -\frac{\tau}{32} Q_2 \left[ \frac{\rho^\sigma}{\pi^\sigma \pi^\text{h} - \left( \pi^\text{x} \right)^2} \right] 
 -\frac{\alpha-\beta}{4\alpha} Q_3 \left[ \frac{\rho^\text{x}}{\pi^\sigma \pi^\text{h} - \left( \pi^\text{x} \right)^2} \right] \notag\\
 &&+\frac{3}{4\alpha}(6-\alpha(4-3\tau)) Q_4 \left[ \frac{\pi^\text{h} \left(\pi^\sigma\rho^\text{h} +
 \pi^\text{h} \rho^\sigma - 2 \pi^\text{x} \rho^\text{x}\right)}{\left(\pi^\sigma \pi^\text{h} - \left( \pi^\text{x} \right)^2\right)^2} \right] 
 +\lambda(1-\tau) Q_3 \left[ \frac{\pi^\text{h} \left(\pi^\sigma\rho^\text{h} +
 \pi^\text{h} \rho^\sigma - 2 \pi^\text{x} \rho^\text{x}\right)}{\left(\pi^\sigma \pi^\text{h} - \left( \pi^\text{x} \right)^2\right)^2} \right] \notag \\
 &&+\frac{\tau}{32} Q_2 \left[ \frac{\pi^\sigma \left(\pi^\sigma\rho^\text{h} +
 \pi^\text{h} \rho^\sigma - 2 \pi^\text{x} \rho^\text{x}\right)}{\left(\pi^\sigma \pi^\text{h} - \left( \pi^\text{x} \right)^2\right)^2} \right] 
 \left.+\frac{\alpha-\beta}{16\alpha} Q_2 \left[ \frac{\pi^\text{x} \left(\pi^\sigma\rho^\text{h} +
 \pi^\text{h} \rho^\sigma - 2 \pi^\text{x} \rho^\text{x}\right)}{\left(\pi^\sigma \pi^\text{h} - \left( \pi^\text{x} \right)^2\right)^2} \right] \right\} \, .
\end{eqnarray}
With field redefinition, define
\begin{align}
 \pi^\sigma_\text{fr} &= -\frac{3}{4\alpha} \left( 4\alpha\lambda(1-\tau) + (\alpha-3)(\Delta+R_k) \right) \\
 \pi^\text{h}_\text{fr} &= -\frac{1}{16\alpha} \left( -4\alpha\lambda(1+\tau) + (3\alpha-\beta^2)(\Delta+R_k) \right) \\
 \pi^\text{x}_\text{fr} &= -\frac{3}{8\alpha}(\alpha-\beta)(\Delta+R_k) \\
 \rho^\sigma_\text{fr} &= \frac{3}{4\alpha}(3-\alpha)(\dot R_k - \eta R_k) \\
 \rho^\text{h}_\text{fr} &= -\frac{1}{16\alpha}(3\alpha-\beta^2)(\dot R_k - \eta R_k) \\
 \rho^\text{x}_\text{fr} &= -\frac{3}{8\alpha}(\alpha-\beta)(\dot R_k - \eta R_k).
\end{align}
Then, the scalar contribution is
\begin{eqnarray}
 S^{\sigma\text{h}}_\text{fr} &=& \frac{1}{2} Q_2 \left[ \frac{\pi^\sigma_\text{fr}\rho^\text{h}_\text{fr} \! +
 \pi^\text{h}_\text{fr} \rho^\sigma_\text{fr} - 2 \pi^\text{x}_\text{fr} \rho^\text{x}_\text{fr}}{\pi^\sigma_\text{fr} \pi^\text{h}_\text{fr} - \left( \pi^\text{x}_\text{fr} \right)^2} \right] 
 +R \left\{ \frac{1}{12} Q_1 \left[ \frac{\pi^\sigma_\text{fr}\rho^\text{h}_\text{fr} +
 \pi^\text{h}_\text{fr} \rho^\sigma_\text{fr} - 2 \pi^\text{x}_\text{fr} \rho^\text{x}_\text{fr}}{\pi^\sigma_\text{fr} \pi^\text{h}_\text{fr} - \left( \pi^\text{x}_\text{fr} \right)^2} \right] \right. 
 -\frac{3}{8\alpha}(1-\alpha(1-\tau)) Q_2 \left[ \frac{\rho^\text{h}_\text{fr}}{\pi^\sigma_\text{fr} \pi^\text{h}_\text{fr} - \left( \pi^\text{x}_\text{fr} \right)^2} \right] \notag \\
 &&-\frac{\tau}{32} Q_2 \left[ \frac{\rho^\sigma_\text{fr}}{\pi^\sigma_\text{fr} \pi^\text{h}_\text{fr} - \left( \pi^\text{x}_\text{fr} \right)^2} \right] 
 -\frac{\alpha-\beta}{16\alpha} Q_2 \left[ \frac{\rho^\text{x}_\text{fr}}{\pi^\sigma_\text{fr} \pi^\text{h}_\text{fr} - \left( \pi^\text{x}_\text{fr} \right)^2} \right] 
 +\frac{3}{8\alpha}(1-\alpha(1-\tau)) Q_2 \left[ \frac{\pi^\text{h}_\text{fr} \left(\pi^\sigma_\text{fr}\rho^\text{h}_\text{fr} +
 \pi^\text{h}_\text{fr} \rho^\sigma_\text{fr} - 2 \pi^\text{x}_\text{fr} \rho^\text{x}_\text{fr}\right)}{\left(\pi^\sigma_\text{fr} \pi^\text{h}_\text{fr} - \left( \pi^\text{x}_\text{fr} \right)^2\right)^2} \right] \notag \\
 &&+\frac{\tau}{32} Q_2 \left[ \frac{\pi^\sigma_\text{fr} \left(\pi^\sigma_\text{fr}\rho^\text{h}_\text{fr} +
 \pi^\text{h}_\text{fr} \rho^\sigma_\text{fr} - 2 \pi^\text{x}_\text{fr} \rho^\text{x}_\text{fr}\right)}{\left(\pi^\sigma_\text{fr} \pi^\text{h}_\text{fr} - \left( \pi^\text{x}_\text{fr} \right)^2\right)^2} \right] 
 \left.+\frac{\alpha-\beta}{16\alpha} Q_2 \left[ \frac{\pi^\text{x}_\text{fr} \left(\pi^\sigma_\text{fr}\rho^\text{h}_\text{fr} +
 \pi^\text{h}_\text{fr} \rho^\sigma_\text{fr} - 2 \pi^\text{x}_\text{fr} \rho^\text{x}_\text{fr}\right)}{\left(\pi^\sigma_\text{fr} \pi^\text{h}_\text{fr} - \left( \pi^\text{x}_\text{fr} \right)^2\right)^2} \right] \right\}  \, .
\end{eqnarray}
\end{widetext}
Further, the ghost contribution reads without field redefinition,
\begin{align}
 S^\text{gh} &= -5 Q_2 \left[ \frac{\dot R_k}{\Delta+R_k} \right] - R \left( \frac{7}{12} Q_1 \left[ \frac{\dot R_k}{\Delta+R_k} \right] \right. \\
 &\left. + \frac{3}{4} Q_2 \left[ \frac{\dot R_k}{(\Delta+R_k)^2} \right] + \frac{4}{3-\beta} Q_3 \left[ \frac{\dot R_k}{(\Delta+R_k)^3} \right] \right) \, . \notag
\end{align}
With field redefinition, it is
\begin{align}
 S^\text{gh}_\text{fr} &= -4 Q_2 \left[ \frac{\dot R_k}{\Delta+R_k} \right] - R \left( \frac{5}{12} Q_1 \left[ \frac{\dot R_k}{\Delta+R_k} \right] \right. \notag \\
 &\left. \qquad + \left( \frac{3}{4} + \frac{1}{3-\beta} \right) Q_2 \left[ \frac{\dot R_k}{(\Delta+R_k)^2} \right] \right) \, .
\end{align}
Finally, the contribution of the Jacobian for the case without field redefinition is
\begin{align}
\mcS^\text{Jac} \! = \! \frac{1}{2} \! \left. S^\text{gh}\right|_{\beta=0} + Q_2 \! \left[ \frac{\dot R_k}{\Delta+R_k} \right] \! + \! \frac{1}{6} R Q_1 \left[ \frac{\dot R_k}{(\Delta+R_k)^2} \right].\label{eq:DefSJac}
\end{align}
With field redefinition, all Jacobians cancel, at least on maximally
symmetric backgrounds, which is sufficient for the truncation
considered here \cite{Lauscher:2001ya}.

\bibliography{general_bib}

\end{document}